\documentclass[prb,aps,12pt]{revtex4-2}
\usepackage{amsmath}
\usepackage{amssymb}
\usepackage{graphicx}

\begin{document}

\title{Macroscopic ground state degeneracy of the ferro-antiferromagnetic
Heisenberg model on diamond-decorated lattices}
\author{D.~V.~Dmitriev}
\email{dmitriev@deom.chph.ras.ru}
\author{V.~Ya.~Krivnov}
\author{O.~A.~Vasilyev}
\affiliation{Institute of Biochemical Physics of RAS, Kosygin str. 4, 119334, Moscow,
Russia.}
\date{}

\begin{abstract}
We investigate the spin-1/2 Heisenberg model with competing
ferromagnetic and antiferromagnetic interactions on
diamond-decorated lattices. Tuning the exchange interactions to
the boundary of the ferromagnetic phase, we analyze the models
with two types of diamond units: distorted and ideal diamonds. In
the distorted diamond model, flat bands in the magnon spectra
indicate the localized states confined to small regions (`trapping
cells') of the lattice. Remarkably, these trapping cells can host
up to five and seven localized states for square and cubic
lattices, respectively, leading to the macroscopic ground state
degeneracy and high value of residual entropy. The problem of
calculating ground state degeneracy reduces to that of
non-interacting spins, whose spin value equal to half the number
of localized magnons in the trapping cell. In contrast, ideal
diamond models feature ground states composed of randomly
distributed isolated diamond diagonal singlets immersed in a
ferromagnetic background. Counting the ground state degeneracies
here maps onto the percolation problem in 2D and 3D lattices. Our
analysis shows that ideal diamond models possess even greater
ground state degeneracy than their distorted counterparts. These
findings suggest that synthesizing diamond-decorated-type
compounds holds great promise for low-temperature cooling
applications.
\end{abstract}

\maketitle

\section{Introduction}

Quantum magnets on geometrically frustrated lattices have been
extensively studied in recent years \cite{diep, diep2, mila}. A
notable class of these systems involves lattices with magnetic
ions located at the vertices of connected triangles. For specific
relations between exchange interactions, these systems exhibit a
dispersionless (flat) one-magnon band. The existence of flat band
in the one-magnon spectrum can be vizualized as the states
localized within a small part of the lattice, `trapping cells',
and it is a consequence of destructive quantum interference. This
phenomenon has been observed in a broad class of highly frustrated
antiferromagnetic spin systems \cite{derzhko2007universal,
schulenburg2002macroscopic, zhitomirsky2005high, flat}. The
localization of one-magnon states forms the basis for constructing
multi-magnon states, because states consisting of isolated
(non-overlapping) localized magnons are exact eigenstates. These
states can be mapped to an effective lattice gas model with a
hard-core potential, enabling the application of classical
statistical mechanics to describe frustrated quantum spin models.
This approach has been widely used for various frustrated quantum
antiferromagnets with flat bands \cite{flat, zhit, zhit2,
strevcka2017diversity, strevcka2022frustrated, caci2023phases,
Derzhko, derzhko2006universal}, including the kagome
antiferromagnet in two dimensions and the pyrochlore
antiferromagnet in three dimensions.

In antiferromagnetic flat-band models, localized states constitute
the ground state manifold in the saturation magnetic field,
leading to an exponentially growing degeneracy in the
thermodynamic limit and residual entropy. The ground state
properties and low-temperature thermodynamics of these models have
been extensively studied, revealing intriguing features such as
zero-temperature magnetization-plateau, an extra low-temperature
peak in the specific heat, and an enhanced magnetocaloric effect
\cite{flat, shulen, zhitomirsky2005high, schmidt, honecker, hon2,
Derzhko, derzhko2006universal, richter2018thermodynamic,
zhitomirsky2003enhanced}.

Another class of frustrated quantum models with an one-magnon flat
band involves systems with competing ferro- and antiferromagnetic
interactions (F-AF models). The zero-temperature phase diagram of
these models exhibits different phases depending on the ratio of
ferromagnetic and antiferromagnetic interactions. At the critical
value of this ratio, corresponding to a phase boundary (quantum
critical point), the model exhibits a macroscopically degenerate
ground state. An example of such F-AF systems is the delta-chain
at the critical value of the frustration parameter \cite{zhitomir,
Derzhko, KDNDR, DKRS, DKRS2, DKRS3} and its 2D generalizations on
Tasaki and Kagome lattices \cite{anis3}. Unlike antiferromagnetic
models, F-AF models feature additional magnon complexes, which are
exact ground states at the critical frustration parameter. This
results in macroscopic ground state degeneracy in zero magnetic
field and a higher residual entropy compared to antiferromagnetic
models. The residual entropy in zero magnetic field enhances
magnetic cooling, which is of practical importance.

Recently, another example of a frustrated F-AF spin model with
macroscopic ground state degeneracy was studied in
\cite{diamond1d}. This model, the spin-(1/2) Heisenberg chain of
distorted diamond units, was shown to exhibit flat bands and a
macroscopically degenerate ground state for specific relations
between exchange interactions \cite{diamond1d}. These conditions
define a critical (transition) point in the parameter space,
marking the transition between ferromagnetic and other (singlet or
ferrimagnetic) ground state phases. Notably, this model features
not only a one-magnon flat band but also two- and three-magnon
dispersionless bands, with the corresponding multi-magnon states
localized in the same trapping cell. All these states belong to
the ground state manifold, leading to an exponential increase in
ground state degeneracy compared to models with only one-magnon
localized states.

Generally, systems consisting of diamond units with frustration
have attracted significant attention both experimentally and
theoretically. One such model is the spin-(1/2) Heisenberg model
on a diamond-decorated square lattice, where the bonds in the
square lattice are replaced by diamonds. A diamond unit with two
different exchange interactions is referred to as an `ideal
diamond', while a unit with three different interactions is called
a `distorted diamond'. Various types of diamond models with
antiferromagnetic exchange interactions have been intensively
studied \cite{morita2016exact, hirose2016exact, hirose2018ground,
caci2023phases, karl2024thermodynamic}. In particular,
one-dimensional and two-dimensional systems composed of ideal
diamond units with antiferromagnetic exchange interactions exhibit
three types of ground state phases, including the Lieb-Mattis
ferrimagnet, a monomer-dimer, and a dimer-tetramer phases. The
latter two phases demonstrate macroscopic ground state degeneracy.

In this paper, we investigate two spin-(1/2) F-AF Heisenberg
models on diamond-decorated two- and three-dimensional lattices.
One model consists of distorted diamond units, while the other
comprises ideal diamonds. We focus on the properties of these
models at the quantum critical line, corresponding to the boundary
between the ferromagnetic and other ground state phases. We
demonstrate that the ground state properties at this line are
highly non-trivial. For instance, the spin-(1/2) F-AF Heisenberg
model on a diamond-decorated square (cubic) lattice with distorted
diamond units hosts up to five (seven) magnon states localized in
the trapping cell, all of which are exact ground states. This
results in macroscopic ground state degeneracy. In contrast, the
ground state of the model with ideal diamonds consists of
ferromagnetic clusters surrounded by regions of diamonds with
diagonal singlets. Counting the number of ground states in this
case reduces to a percolation problem, and the ground state
degeneracy of the ideal diamond model exceeds that of the
distorted diamond model.

The paper is organized as follows. In Sec. II, we introduce the
Hamiltonian of the spin-$\frac{1}{2}$ F-AF Heisenberg model with
distorted diamonds and identify the conditions on the exchange
interactions that correspond to the critical line. We demonstrate
that for any lattice or finite graph composed of distorted
diamonds connected by ferromagnetic bonds, trapping cells can host
several localized magnons, contributing to the macroscopic
degeneracy of the ground state. Importantly, this ground state
manifold is equivalent to that of non-interacting spins with
magnitudes $\frac{z+1}{2}$, where $z$ represents the coordination
number of the lattice. In Sec. III, we turn our attention to
models featuring ideal diamonds. Here, we calculate the ground
state degeneracy and magnetization by mapping the problem to a
percolation framework. Finally, in the concluding Section, we
summarize our key findings.

\section{F-AF Heisenberg model on the distorted diamond decorated
lattices}

\begin{figure}[tbp]
\includegraphics[width=5in,angle=0]{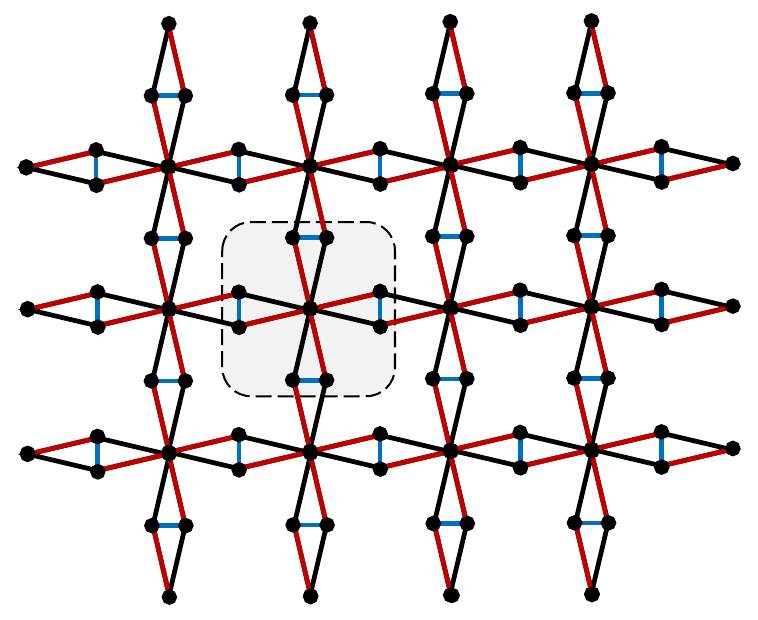}
\caption{The spin-1/2 Heisenberg model on the diamond decorated square
lattice. Shaded area indicates the trapping star.}
\label{Fig_square_lattice}
\end{figure}

In this section, we will study the spin-$\frac{1}{2}$ F-AF Heisenberg model
on 2D and 3D distorted diamond-decorated lattices. We will show that the
ground state is macroscopically degenerate on the phase boundary between the
ferromagnetic and singlet ground states. For definiteness we consider the
diamond-decorated square lattice shown in Fig.\ref{Fig_square_lattice}.
However, all the analysis and the obtained results directly apply to any 2D
and 3D lattices.

The Hamiltonian of the diamond-decorated square model can be written as a
sum of diamond Hamiltonians
\begin{equation}
\hat{H}=\sum_{\left\langle \mathbf{i,j}\right\rangle }\hat{H}_{\mathbf{i,j}}
\label{H0}
\end{equation}%
where the sum is taken over all diamonds of the system, located
between two neighboring nodes $\mathbf{i=}\left(
i_{x},i_{y}\right) $ and $\mathbf{j=}\left( j_{x},j_{y}\right) $
of the square lattice.

\begin{figure}[tbp]
\includegraphics[width=3in,angle=0]{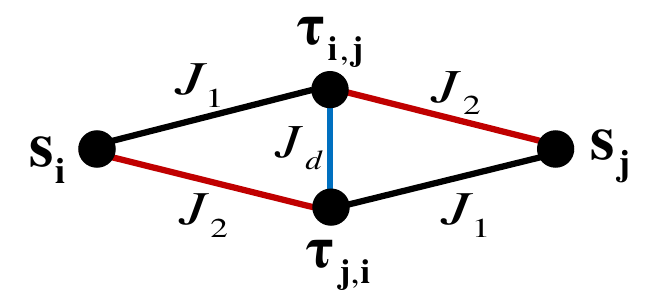}
\caption{Distorted diamond. Exchange interactions $J_1$, $J_2$ and $J_d$ are
shown by black, red and blue lines, respectively.}
\label{Fig_diamond}
\end{figure}

The distorted diamond formed by two central spins
$\mathbf{s}_{\mathbf{i}}$ and $\mathbf{s}_{\mathbf{j}}$\ and two
diagonal spins $\mathbf{\tau }_{\mathbf{i},\mathbf{j}}$ and
$\mathbf{\tau }_{\mathbf{j,i}}$ is shown in Fig.\ref{Fig_diamond}.
According to Fig.\ref{Fig_diamond}, the Hamiltonian of this
diamond has the form
\begin{equation}
\hat{H}_{\mathbf{i},\mathbf{j}}=J_{1}\left( \mathbf{s}_{\mathbf{i}}\cdot
\mathbf{\tau }_{\mathbf{i},\mathbf{j}}\mathbf{+s}_{\mathbf{j}}\cdot \mathbf{%
\tau }_{\mathbf{j,i}}\right) \mathbf{+}J_{2}\left( \mathbf{s}_{\mathbf{i}%
}\cdot \mathbf{\tau }_{\mathbf{j,i}}\mathbf{+s}_{\mathbf{j}}\cdot \mathbf{%
\tau }_{\mathbf{i},\mathbf{j}}\right) \mathbf{+}J_{d}\mathbf{\tau }_{\mathbf{%
i},\mathbf{j}}\cdot \mathbf{\tau }_{\mathbf{j,i}}-J_{0}  \label{Hd}
\end{equation}%
where
$\mathbf{s}_{\mathbf{i}},\mathbf{s}_{\mathbf{j}},\mathbf{\tau
}_{\mathbf{i},\mathbf{j}}$\ and $\mathbf{\tau }_{\mathbf{j,i}}$\
are spin-$1/2$\ operators and the constant
$J_{0}=\frac{1}{4}(J_{d}+2J_{1}+2J_{2})$ in (\ref{Hd}) is chosen
so that the energy of the ferromagnetic state of diamond ($S=2$)
is zero. We study the model on the phase boundary between the
ferromagnetic and singlet ground states, which requires that one
of side exchange interactions $J_{1}$ or/and $J_{2}$ be
ferromagnetic. We assume that $J_{1}\leq J_{2}$, so that $J_{1}$
is the strongest ferromagnetic interaction, which also scales the
energy. Thus, we set $J_{1}=-1$ and the interaction $J_{2}=J$ can
be both ferromagnetic or antiferromagnetic.

At first, we need to determine the condition, for which model (\ref{H0}) is
on the phase boundary between the ferromagnetic and singlet ground state. As
it was shown in \cite{diamond1d} this condition imposes the following
restrictions on the exchange integrals:%
\begin{equation}
J_{d}=\frac{2J}{J-1}  \label{cond1}
\end{equation}%
and%
\begin{equation}
-1<J<1  \label{cond2}
\end{equation}

For $J<-1$ the model can be reduced to the same range by replacing the
diamond sides and rescaling the energy. For $J=-1$ the distorted diamond
model becomes the ideal diamond model, which will be studied in Section 3.
In the case of $J>1$, the ferromagnetic state is no longer the ground state.

Conditions (\ref{cond1}) and (\ref{cond2}) define the transition line on the
ground state phase diagram in the ($J$, $J_{d}$) plane. As it will be shown
below the 2D and 3D distorted diamond models on the transition line have
exact localized magnon states and the macroscopic degeneracy of the ground
state.

Let us analyze the Hamiltonian of the diamond unit (\ref{Hd}).
Generally, four spins in the diamond unit form one quintet
($S=2$), three triplets ($S=1 $) and two singlets ($S=0$). The
condition (\ref{cond1}) secures that the quintet, one of three
triplets and one of two singlets of
$\hat{H}_{\mathbf{i},\mathbf{j}}$ (\ref{Hd}) are degenerate with
$E=0$. The condition (\ref{cond2}) provides that these nine
degenerate states are ground states of
$\hat{H}_{\mathbf{i},\mathbf{j}}$, so that all other eigenvalues
$E_{i}>0$. Thus, the conditions for the transition line
automatically leads to nine-fold degeneracy of the ground state
for the distorted diamond unit. Further we assume that both
conditions (\ref{cond1}) and (\ref{cond2}) are always satisfied.

Now we prove that the energy of the ground state of the total
Hamiltonian is zero. Because the neighboring diamond unit
Hamiltonians $\hat{H}_{\mathbf{i},\mathbf{j}}$ and
$\hat{H}_{\mathbf{j},\mathbf{k}}$ do not commute with each other,
the following inequality for the lowest eigenvalue $E_{0}$ of
$\hat{H}
$ is valid:%
\begin{equation}
E_{0}\geq \sum E_{i}=0  \label{ineq}
\end{equation}

The energy of the ferromagnetic state of $\hat{H}$ with maximal
total spin $S_{\max }=\frac{\mathcal{N}}{2}$ is zero. (Hereinafter
we use the following notations: $N$ is the number of central spins
$\mathbf{s}_{\mathbf{i}}$, $N_{b}$ is the number of diamonds and
$\mathcal{N}$\ is total number of spins in the system.) Therefore,
the inequality in (\ref{ineq}) turns into an equality and the
ground state energy of $\hat{H}$ is zero.

\subsection{Multi-magnon localized states in a trapping cell}

In this subsection, we will construct multi-magnon states localized in one
trapping cell and prove that the constructed states are the exact ground
states of the model. For the construction of localized multi-magnon states,
it is useful to write down the following linear combinations of nine ground
state functions of $\hat{H}_{\mathbf{i,j}}$:%
\begin{eqnarray}
&&\left( s_{\mathbf{i}}^{+}+\sigma _{\mathbf{i},\mathbf{j}}^{+}\right)
\left\vert F\right\rangle  \label{1} \\
&&s_{\mathbf{i}}^{+}\sigma _{\mathbf{i},\mathbf{j}}^{+}\left\vert
F\right\rangle  \label{2} \\
&&\left( s_{\mathbf{j}}^{+}+\sigma _{\mathbf{j,i}}^{+}\right) \left\vert
F\right\rangle  \label{3} \\
&&s_{\mathbf{j}}^{+}\sigma _{\mathbf{j,i}}^{+}\left\vert F\right\rangle
\label{4}
\end{eqnarray}%
where $\left\vert F\right\rangle $ is fully polarized state with all spins
down and $s_{\mathbf{i}}^{+}$ are raising spin operators. Here we also
introduced two new convenient operators on diagonal of each diamond:
\begin{eqnarray}
\sigma _{\mathbf{i},\mathbf{j}}^{+} &=&\frac{\tau _{\mathbf{i},\mathbf{j}%
}^{+}+J\tau _{\mathbf{j,i}}^{+}}{1+J} \\
\sigma _{\mathbf{j,i}}^{+} &=&\frac{\tau _{\mathbf{j,i}}^{+}+J\tau _{\mathbf{%
i},\mathbf{j}}^{+}}{1+J}
\end{eqnarray}%
so that%
\begin{equation}
\sigma _{\mathbf{i},\mathbf{j}}^{+}+\sigma _{\mathbf{j,i}}^{+}=\tau _{%
\mathbf{i},\mathbf{j}}^{+}+\tau _{\mathbf{j,i}}^{+}
\end{equation}

One can directly verify that four states (\ref{1})-(\ref{4}) are ground
states of $\hat{H}_{\mathbf{i,j}}$ (\ref{Hd}), though they are not
eigenstates of the total spin operator of diamond.

\begin{figure}[tbp]
\includegraphics[width=3in,angle=0]{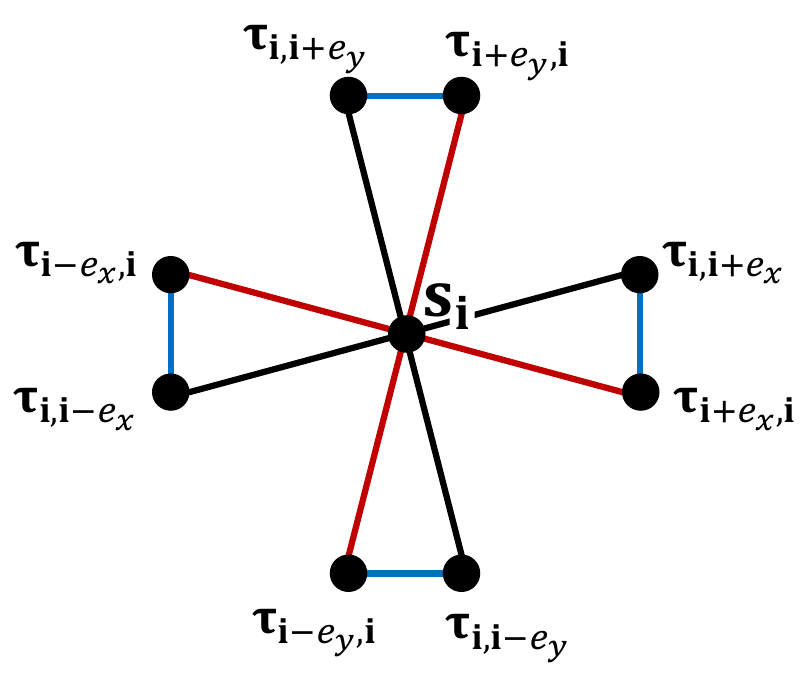}
\caption{Spin-star cluster on the distorted diamond-decorated square lattice.
}
\label{Fig_star}
\end{figure}

As will be shown below, the multi-magnon states are localized in the
trapping cells, one of which is indicated in Fig.\ref{Fig_square_lattice} by
a shaded region. Therefore, it is convenient to rewrite the Hamiltonian of
the diamond-decorated square model (\ref{H0}) as a sum of trapping star
cluster Hamiltonians
\begin{equation}
\hat{H}=\sum_{\mathbf{i}}\hat{H}_{\mathbf{i}}  \label{H}
\end{equation}%
where trapping spin star cluster is shown in Fig.\ref{Fig_star},
$\mathbf{i}=\left( i_{x},i_{y}\right) $ numbers central spins
$\mathbf{s}_{\mathbf{i}}$ and the corresponding spin star cluster.

The diamond-decorated square lattice consists of $N=N_{x}N_{y}$ spin star
clusters (with total number of spins $\mathcal{N}=5N$) and periodical
boundary conditions are assumed. The Hamiltonian of $\mathbf{i}$-th trapping
star cluster has the form%
\begin{equation}
\hat{H}_{\mathbf{i}}=J_{1}\sum_{\delta =\pm e_{x},\pm e_{y}}\mathbf{s}_{%
\mathbf{i}}\cdot \mathbf{\tau }_{\mathbf{i},\mathbf{i}+\delta
}+J_{2}\sum_{\delta =\pm e_{x},\pm e_{y}}\mathbf{s}_{\mathbf{i}}\cdot
\mathbf{\tau }_{\mathbf{i}+\delta ,\mathbf{i}}+\frac{J_{d}}{2}\sum_{\delta
=\pm e_{x},\pm e_{y}}\mathbf{\tau }_{\mathbf{i},\mathbf{i}+\delta }\cdot
\mathbf{\tau }_{\mathbf{i}+\delta ,\mathbf{i}}+J_{0}  \label{Hi}
\end{equation}%
where $e_{x}$ and $e_{y}$ are unit vectors in $X$ and $Y$ directions. The
third term in Eq.(\ref{Hi}) has the factor $\frac{1}{2}$, because this term
arises in the sum (\ref{H}) twice.

The general form of the $k$-magnon state localized in the trapping
star $\mathbf{i}$ can be written as
$\hat{\varphi}_{\mathbf{i}}^{(k)}\left\vert F\right\rangle $ with:
\begin{equation}
\hat{\varphi}_{\mathbf{i}}^{(k)}=s_{\mathbf{i}}^{+}\sum_{\left\{ \mathbf{r}%
_{1}\ldots \mathbf{r}_{k-1}\right\} }\sigma _{\mathbf{i},\mathbf{r}%
_{1}}^{+}\ldots \sigma _{\mathbf{i},\mathbf{r}_{k-1}}^{+}+\sum_{\left\{
\mathbf{r}_{1}\ldots \mathbf{r}_{k}\right\} }\sigma _{\mathbf{i},\mathbf{r}%
_{1}}^{+}\ldots \sigma _{\mathbf{i},\mathbf{r}_{k}}^{+}  \label{phi_k}
\end{equation}

Here the first sum is taken over all configurations $\left\{
\mathbf{r}_{1}\ldots \mathbf{r}_{k-1}\right\} $ under the
constraint $\mathbf{r}_{i}\neq \mathbf{r}_{j}$ and the second sum
over all configurations $\left\{ \mathbf{r}_{1}\ldots
\mathbf{r}_{k}\right\} $ with the same constraint
$\mathbf{r}_{i}\neq \mathbf{r}_{j}$.

The explicit form of one and two-magnon functions for 2D square lattice are:%
\begin{eqnarray}
\hat{\varphi}_{\mathbf{i}}^{(1)} &=&s_{\mathbf{i}}^{+}+\sum_{\delta =\pm
e_{x},\pm e_{y}}\sigma _{\mathbf{i},\mathbf{i}+\delta }^{+}  \label{phi_1} \\
\hat{\varphi}_{\mathbf{i}}^{(2)} &=&s_{\mathbf{i}}^{+}\sum_{\delta =\pm
e_{x},\pm e_{y}}\sigma _{\mathbf{i},\mathbf{i}+\delta }^{+}+\sum_{\substack{ %
\delta _{1},\delta _{2}=\pm e_{x},\pm e_{y}  \\ \delta _{1}\neq \delta _{2}}}%
\sigma _{\mathbf{i},\mathbf{i}+\delta _{1}}^{+}\sigma _{\mathbf{i},\mathbf{i}%
+\delta _{2}}^{+}  \label{phi_2}
\end{eqnarray}

The number of adjacent diamonds to the $\mathbf{i}$-th central
spin is four, therefore the number of different operators $\sigma
_{\mathbf{i},\mathbf{r}_{1}}^{+}$ is also four. Hence, as follows
from Eq.(\ref{phi_k}), the maximal number of magnons located in
the $\mathbf{i}$-th trapping star cluster is five, and the
corresponding function is simple:\begin{equation}
\hat{\varphi}_{\mathbf{i}}^{(5)}=s_{\mathbf{i}}^{+}\sigma _{\mathbf{i},%
\mathbf{i}+e_{x}}^{+}\sigma _{\mathbf{i},\mathbf{i}-e_{x}}^{+}\sigma _{%
\mathbf{i},\mathbf{i}+e_{y}}^{+}\sigma _{\mathbf{i},\mathbf{i}-e_{y}}^{+}
\end{equation}

In general, the maximum number of localized magnons $k_{\max }$ in a single
trap star cluster (\ref{phi_k}) depends only on the coordination number $z$
of the lattice under study, as follows

\begin{equation}
k_{\max }=z+1  \label{kmax}
\end{equation}

The proof that the states (\ref{phi_k}) are exact ground state of the total
Hamiltonian (\ref{H}) is given in Appendix A.

\begin{figure}[tbp]
\includegraphics[width=5in,angle=0]{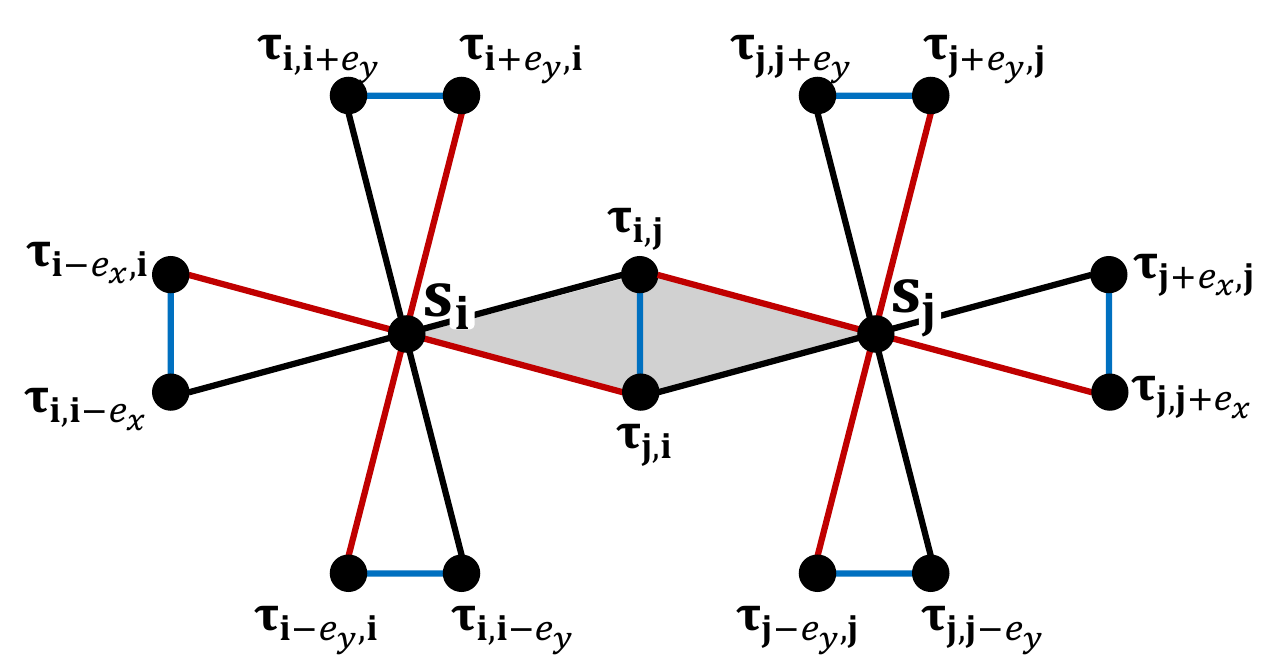}
\caption{Two-neighboring star clusters. Distorted diamond
$(\mathbf{i},\mathbf{j})$ between stars is shaded.}
\label{Fig_star2}
\end{figure}

The localized magnons $\hat{\varphi}_{\mathbf{i}}^{(k)}\left\vert
F\right\rangle $ affect the spins in the $\mathbf{i}$-th star
only. Therefore, if we place several localized magnons in the
$\mathbf{i}$-th star and several in the $\mathbf{j}$-th star,
provided that the $\mathbf{i}$-th and $\mathbf{j}$-th stars are
not neighbors, the resulting state
$\hat{\varphi}_{\mathbf{i}}^{(1)}\hat{\varphi}_{\mathbf{j}}^{(1)}\left\vert
F\right\rangle $ will be the exact ground state of the total
Hamiltonian $\hat{H}$.

The problem of calculating the degeneracy of the ground state
based on the constructed functions
$\hat{\varphi}_{\mathbf{i}}^{(k)}\left\vert F\right\rangle $ lies
in the fact that localized magnon states (\ref{phi_k}) cannot be
located in neighboring stars. For example, let us consider a
diamond located between two adjacent stars (see
Fig.\ref{Fig_star2}). The product of two one-magnon functions
located in these stars
$\hat{\varphi}_{\mathbf{i}}^{(1)}\hat{\varphi}_{\mathbf{j}}^{(1)}\left\vert
F\right\rangle $, for example, is not the exact state of the
Hamiltonian $\hat{H}_{\mathbf{i},\mathbf{j}}$ (\ref{Hd}).
Therefore, in order to calculate the ground state degeneracy of
the system, it is necessary to take into account the restriction
that localized multi-magnon functions cannot touch each other.
However, as will be shown in the next Section, the above
restriction on neighboring localized magnons can be lifted, which
significantly simplifies the calculation of the ground state
degeneracy.

\subsection{Correction terms for neighboring localized magnon states}

In this subsection, we present a method for removing the
restriction on neighboring localized magnons. First, let us
consider a diamond $(\mathbf{i},\mathbf{j})$ located in two
adjacent stars, shown in Fig.\ref{Fig_star2}. As was noted in the
previous Section, the product of one-magnon functions located in
the neighboring stars
$\hat{\varphi}_{\mathbf{i}}^{(1)}\hat{\varphi}_{\mathbf{j}}^{(1)}\left\vert
F\right\rangle $ is not the exact state of the Hamiltonian
$\hat{H}_{\mathbf{i},\mathbf{j}}$ (\ref{Hd}) of the diamond
($i,j)$. However, the exact ground state can be constructed from
the state
$\hat{\varphi}_{\mathbf{i}}^{(1)}\hat{\varphi}_{\mathbf{j}}^{(1)}\left\vert
F\right\rangle $ by adding to it a certain correction term:
\begin{equation}
\hat{\varphi}_{\mathbf{i}}^{(1)}\hat{\varphi}_{\mathbf{j}}^{(1)}\left\vert
F\right\rangle +\frac{2J}{(1+J)^{2}}\tau _{\mathbf{i},\mathbf{j}}^{+}\tau _{%
\mathbf{j,i}}^{+}\left\vert F\right\rangle  \label{correction}
\end{equation}

One can directly check that the function (\ref{correction}) is
exact ground state for the Hamiltonian
$\hat{H}_{\mathbf{i},\mathbf{j}}$ (\ref{Hd}) and for Hamiltonians
of all other adjacent diamonds $\hat{H}_{\mathbf{i},\mathbf{k}}$
and $\hat{H}_{\mathbf{j},\mathbf{k}}$.

This means that two-magnon sector of the ground state manifold of
the diamond-decorated square lattice with $N$\ central spins
contains: $N$ states with two magnons in the same star,
$\hat{\varphi}_{\mathbf{i}}^{(2)}\left\vert F\right\rangle $;
$N(N-5)/2$ isolated magnons states with non-neighboring $i$ and
$j$,
$\hat{\varphi}_{\mathbf{i}}^{(1)}\hat{\varphi}_{\mathbf{j}}^{(1)}\left\vert
F\right\rangle $; and $2N$ states like (\ref{correction}) with
magnons located in neighboring stars, but with correction term.
So, it's as if the restriction on neighboring one-magnon states
has been lifted.

Generally, the state with $k$-magnons in the left star and
$m$-magnons in the right star,
$\hat{\varphi}_{\mathbf{i}}^{(k)}\hat{\varphi}_{\mathbf{j}}^{(m)}$
(see Fig.\ref{Fig_star2}), is not exact for the Hamiltonian
$\hat{H}_{\mathbf{i},\mathbf{j}}$ (\ref{Hd}). However, one can
construct the exact ground state be adding the following
correction term:
\begin{equation}
\hat{\varphi}_{\mathbf{i}}^{(k)}\hat{\varphi}_{\mathbf{j}}^{(m)}\left\vert
F\right\rangle +\frac{2J}{(1+J)^{2}}\tau _{\mathbf{i},\mathbf{j}}^{+}\tau _{%
\mathbf{j,i}}^{+}\hat{\varphi}_{\mathbf{i}}^{(k-1)}\hat{\varphi}_{\mathbf{j}%
}^{(m-1)}\left\vert F\right\rangle  \label{corr-term}
\end{equation}

For further analysis, it is convenient to introduce the correction
operator $\hat{\delta}_{\mathbf{i},\mathbf{j}}$, which is defined
as
\begin{equation}
\hat{\delta}_{\mathbf{i},\mathbf{j}}\hat{\varphi}_{\mathbf{i}}^{(k)}\hat{%
\varphi}_{\mathbf{j}}^{(m)}=\frac{2J}{(1+J)^{2}}\tau _{\mathbf{i},\mathbf{j}%
}^{+}\tau _{\mathbf{j,i}}^{+}\hat{\varphi}_{\mathbf{i}}^{(k-1)}\hat{\varphi}%
_{\mathbf{j}}^{(m-1)}  \label{delta}
\end{equation}

We notice that $\hat{\varphi}_{\mathbf{i}}^{(0)}=1$ and the
correction term (\ref{corr-term}) reduces to that in
Eq.(\ref{correction}) for $k=m=1$.

The construction of the states with correction terms effectively
removes the restriction on the neighboring localized magnon
states. Thus, we developed the method for constructing the ground
states based on magnons localized in two neighboring stars,
provided that all other stars are empty. Now we need to generalize
the method for the case when all stars are occupied by different
numbers of localized magnons. In other words, we should construct
the ground state $\Psi $, corresponding to the configuration with
$k_{\mathbf{i}}$ magnons in the $\mathbf{i}$-th stars, $\left\{
k_{\mathbf{i} }\right\} $:
\begin{equation}
\Phi \left( \left\{ k_{\mathbf{i}}\right\} \right) =\prod_{\mathbf{i}}\hat{%
\varphi}_{\mathbf{i}}^{(k_{\mathbf{i}})}\left\vert F\right\rangle
\label{Phi}
\end{equation}

The wave function $\Phi $ (\ref{Phi}) is not exact state of
Hamiltonian (\ref{H}). In order to correct $\Phi $ we add the
correction terms $\hat{\delta}_{\mathbf{j},\mathbf{m}}$ for all
pairs of neighboring stars, so that the exact ground state is
written as
\begin{equation}
\Psi \left( \left\{ k_{\mathbf{i}}\right\} \right) =\prod_{\left\langle
\mathbf{j},\mathbf{m}\right\rangle }\left( 1+\hat{\delta}_{\mathbf{j},%
\mathbf{m}}\right) \Phi \left( \left\{ k_{\mathbf{i}}\right\} \right)
\label{Psi}
\end{equation}

In order to verify that $\Psi $ is exact ground state we check
whether $\Psi$ is exact for one diamond Hamiltonian, say
$\hat{H}_{\mathbf{1},\mathbf{2}}$ (\ref{Hd}). To do this, we
extract the multiplier $(1+\hat{\delta}_{1,2})$ from the product
and represent (\ref{Psi}) as
\begin{equation}
\Psi \left( \left\{ k_{\mathbf{i}}\right\} \right) =\left( 1+\hat{\delta}%
_{1,2}\right) \prod_{\left\langle \mathbf{j},\mathbf{m}\right\rangle
}^{\prime }\left( 1+\hat{\delta}_{\mathbf{j},\mathbf{m}}\right) \Phi \left(
\left\{ k_{\mathbf{i}}\right\} \right)
\end{equation}

Here the product is taken over all neighboring pair $\left\langle
\mathbf{j},\mathbf{m}\right\rangle $, with the exception of the
pair $\left\langle 1,2\right\rangle $. The wave function
\begin{equation}
\prod_{\left\langle \mathbf{j},\mathbf{m}\right\rangle }^{\prime }\left( 1+%
\hat{\delta}_{\mathbf{j},\mathbf{m}}\right) \Phi \left( \left\{ k_{\mathbf{i}%
}\right\} \right)
\end{equation}%
contains a lot of terms, but they all contain a multiplier related
to stars $1$ and $2$, only in the form of
$\hat{\varphi}_{\mathbf{1}}^{(k)}\hat{\varphi}_{\mathbf{2}}^{(m)}$
with various $k$ and $m$ in different terms. But the function
$(1+\hat{\delta}_{1,2})\hat{\varphi}_{\mathbf{1}}^{(k)}\hat{\varphi}_{\mathbf{2}}^{(m)}$
with any $k$ and $m$ is exact ground state for the Hamiltonian
$\hat{H}_{\mathbf{1},\mathbf{2}}$. Therefore, $\Psi $ is exact
ground state of $\hat{H}_{\mathbf{1},\mathbf{2}}$. Repeating the
same arguments for all other $\hat{H}_{\mathbf{i},\mathbf{j}}$, we
conclude that $\Psi $ is exact ground state of the total
Hamiltonian (\ref{H0}).

Thus, for any given configuration with $k_{\mathbf{i}}$ magnons in
the $\mathbf{i}$-th stars, $\left\{ k_{\mathbf{i}}\right\} $, it
is possible to construct the ground state (with the correction
factors) including states with magnons on neighboring trapping
stars. As will be shown in the next Section this statement
significantly simplifies the calculation of the ground state
degeneracy.

\subsection{Total ground state degeneracy}

The form of localized multi-magnon functions (\ref{phi_k}) and the
procedure for constructing of the ground state (\ref{Psi})
corresponding to any configuration with $k_{\mathbf{i}}$ magnons
in the $\mathbf{i}$-th stars, $\left\{ k_{\mathbf{i}}\right\} $,
(including configurations on neighboring stars) are valid for any
lattice: 1D chain, 2D square, hexagonal or triangle lattices, 3D
cubic etc. The only difference between the different lattices is
the different maximum number of magnons that can be located in
each star.

For the beginning we calculate the ground state degeneracy of the
spin-$\frac{1}{2}$ F-AF Heisenberg model (\ref{H0}) on the
diamond-decorated square lattice. As it was mentioned, each spin
star cluster (Fig.\ref{Fig_star}) can contain up to five localized
magnons presented in Eqs.(\ref{phi_k}) and there is no other
restrictions on possible configurations with
$k_{\mathbf{i}}=0,1\ldots 5$ magnons in the $\mathbf{i}$-th stars,
$\left\{ k_{\mathbf{i}}\right\} $. Since the configuration with
all $k_{\mathbf{i}}=0$ corresponds to a fully polarized
ferromagnetic state $\left\vert F\right\rangle $ with all spins
down, a correspondence can be drawn between the configurations
$\left\{ k_{\mathbf{i}}\right\} $ and the system of $n$
independent spins-$\frac{5}{2}$: the states with
$k_{\mathbf{i}}=0,1,2,3,4,5$ magnons in the $\mathbf{i}$-th stars
corresponds to the states with
$S^{z}=-\frac{5}{2},-\frac{3}{2},-\frac{1}{2},\frac{1}{2},\frac{3}{2},\frac{5}{2}$
of the $\mathbf{i}$-th effective spin-$\frac{5}{2}$. Therefore,
the problem of the ground state degeneracy on the square lattice
is reduced to counting the number of states with given $S^{z}$
for\ the system of $N$\ non-interacting spins-$\frac{5}{2}$. In
particular, the total number of ground states $W(N)$ of the F-AF
spin-$\frac{1}{2}$ Heisenberg model on the diamond-decorated
square lattice is
\begin{equation}
W(N)=6^{N}
\end{equation}%
and the residual entropy per spin $\mathcal{S}_{0}=\ln W/N$ is%
\begin{equation}
\mathcal{S}_{0}=\frac{1}{5}\ln 6=0.3584
\end{equation}%
which is $51\%$ of the maximal value $\mathcal{S}_{\max }\mathcal{=}\ln 2$.

Similar analysis can be extended to the spin-$\frac{1}{2}$ F-AF Heisenberg
model on the diamond - decorated cubic lattice, for which the corresponding
spin star cluster can contain up to seven localized magnons. The number of
the ground states in the spin sector $S^{z}=\frac{7N}{2}-k$ is equal to the
number of the corresponding states in the system of $N$ non-interacting
spins $\frac{7}{2}$. The total number of the ground states of the F-AF model
on diamond-decorated cubic lattice is
\begin{equation}
W(N)=8^{N}
\end{equation}%
and the residual entropy is
\begin{equation}
\mathcal{S}_{0}=\frac{1}{7}\ln 8=0.297
\end{equation}

In general, the ground state degeneracy of the spin-$\frac{1}{2}$
F-AF Heisenberg model can be calculated for any diamond-decorated
lattice. For the lattice with coordination number $z$ the ground
state manifold is equivalent to the manifold of $N$
non-interacting spins with spin value $\frac{z+1}{2}$ and the
ground state degeneracy is
\begin{equation}
W(N)=(z+2)^{N}
\end{equation}%
and the corresponding residual entropy is
\begin{equation}
\mathcal{S}_{0}=\frac{1}{z+1}\ln (z+2)
\end{equation}

For example, $W(N)=5^{N}$ and $W(N)=8^{N}$ for the 2D hexagonal and the
triangle lattices, respectively.

\subsection{Generalizations to other lattices and graphs}

All the above results are valid for the F-AF distorted diamond
model if the interaction $J$ is in range $-1<J<1$. For $J=-1$ the
distorted diamond model becomes the ideal diamond model, which
will be studied in Section 3. In the case of $J>1$, the inequality
(\ref{cond2}) does not hold and the ground state is no longer
macroscopically degenerate. For the case $J\to 1$ the diagonal
interaction (\ref{cond1}) $J_{d}\rightarrow \infty $ and the model
effectively transforms to the trivial model with non-interacting
ferromagnetic diagonals and free central spins, in which the
ground state degeneracy is $W(N)=18^{N}$ and $W(N)=54^{N}$ for the
square and cubic lattices, respectively.

\begin{figure}[tbp]
\includegraphics[width=5in,angle=0]{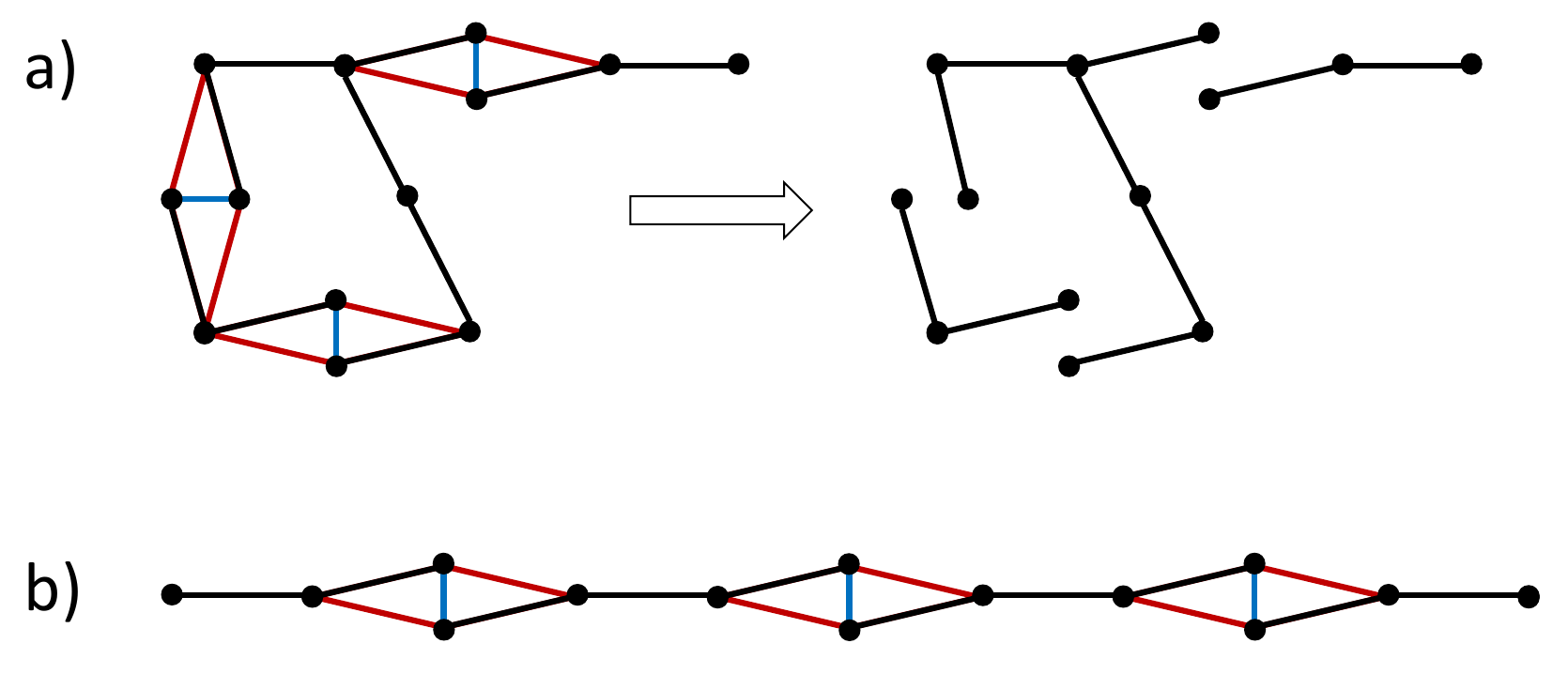}
\caption{ a) Example of graph: three distorted diamonds connected by
ferromagnetic interactions. b) Chain of distorted diamonds connected by
ferromagnetic interactions.}
\label{Fig_graph}
\end{figure}

The most important case is the case of $J=0$, which seems trivial
at first glance, but it turns out to be generic. In this case, all
interactions $J$ and $J_{d}$ (red and green bonds in
Fig.\ref{Fig_square_lattice}) disappear and the square
diamond-decorated model decomposes into $N$ independent five-spin
ferromagnetic clusters. The ground state of each cluster is
ferromagnetic with the total spin $\frac{5}{2}$, which immediately
produces the ground state degeneracy $W(N)=6^{N}$, as for the case
$J\neq 0$. Thus, the ground state degeneracy of the independent
clusters is not lifted by the non-zero exchange interaction $J$.
This remarkable property holds for different infinite and finite
lattices or even graphs, decorated with distorted diamonds. Some
examples of such systems are shown in Fig.\ref{Fig_graph}. The
Hamiltonian of these systems can be written as
\begin{equation}
\hat{H}=\hat{H}_{0}+\hat{V}
\end{equation}%
where $\hat{H}_{0}$ is the Hamiltonian of the model with
independent ferromagnetic clusters of any forms (at $J=0$) and
$\hat{V}$ includes all interaction $J$ and $J_{d}$. The ground
state degeneracy of both models with $\hat{H}_{0}$ and $\hat{H}$
is the same. This amazing feature can be used to construct a
mnemonic rule for calculating the ground state degeneracy of any
lattice or graph: set all the interactions $J$ and $J_{d}$ equal
to zero, and calculate the ground state degeneracy of the system
of the resulting independent ferromagnetic clusters. For example,
the graph shown in Fig.\ref{Fig_graph}a consists of three diamonds
connected by the ferromagnetic interactions, and at $J=0$ it
decomposes into three ferromagnetic clusters with the total spin
of ground states $\frac{3}{2},\frac{7}{2},\frac{3}{2}$. The ground
state degeneracy of the system of three independent spins
$\frac{3}{2},\frac{7}{2},\frac{3}{2}$ is $W(N)=4\cdot 8\cdot
4=128$, and this degeneracy remains for any parameter $J<1$.
Following the above mnemonic rule, the diamond chain, shown in
Fig.\ref{Fig_graph}b, decomposes at $J=0$ into $N$ four-spin
ferromagnetic clusters, and the total ground state degeneracy is
$W(N)=5^{N}$ for both cases $J=0$ and $J\neq 0$.

The class of the diamond-decorated models with macroscopic degeneracy of the
ground state can be extended to models in which all diamonds in the system
have different interactions $J_{i}$ (and corresponding $J_{d,i}$), as well
as to any values of the central spins $s_{i}$.

The generic case ($J=0$) also allows us to determine the nature of
the phase transition described by the model (\ref{H0}). For the
lattice with coordination number $z$ in the generic case ($J=0$)
the system decomposes into the independent ferromagnetic clusters
containing $(z+1)$ spins. When $J$ is small, these ferromagnetic
clusters weakly interact with each other by two types of
interactions $J$ and $J_{d}$. On the transition line
$J_{d}=2J/(J-1)$ (\ref{cond1}), the exchange interactions $J$ and
$J_{d}$ have different signs and compensate each other, so that
the ground state remains macroscopically degenerate including
states with all possible values of total spin from $S=0$ to
$S=S_{\max }$. When the relation between $J$ and $J_{d}$ deviates
from Eq.(\ref{cond1}), the model reduces to the system of
effective spins $s=\frac{z+1}{2}$ on the corresponding lattice,
weakly interacting with each other. When $J_{d}<2J/(J-1)$, the net
interaction is of ferromagnetic type and the ground state is
ferromagnetic. When $J_{d}>2J/(J-1)$, the net interaction is of
antiferromagnetic type and the system reduces to the
antiferromagnetic model of effective spins $s=\frac{z+1}{2}$ on
the corresponding lattice. The ground state of such system is
singlet. Thus, in the vicinity of the point $J=0$, the line
(\ref{cond1}) defines the transition between the ferromagnetic and
singlet ground states. We expect this type of phase transition to
extend to the entire transition line (\ref{cond1}) in the region
$-1<J<1$.

\subsection{Energy gap}

As was shown in previous subsections, the ground state manifold of
the distorted diamond decorated models satisfying the conditions
(\ref{cond1}) and (\ref{cond2}) is the same as that for the system
of independent spins, the value of which relates to the
coordination number of the lattice $z$. This means that the
low-temperature magnetic properties of the complicated frustrated
spin system can be described by that of just one spin
$s=\frac{z+1}{2}$. This means that the low-temperature magnetic
properties of a complex frustrated spin system can be described by
the behavior of only one spin $s=\frac{z+1}{2}$ in a magnetic
field. The question is, how long will this simple picture last
when the temperature rises? To answer this question, we need to
study the energy spectrum of the model. As was shown in
\cite{diamond1d}, there is a finite excitation gap in the 1D chain
of distorted diamond. Here we study the excitation spectrum of the
diamond decorated square lattice, shown in
Fig.\ref{Fig_square_lattice}.

At first, we analyze the one-magnon spectrum. In the spin sector
$S^{z}=S_{\max }-1$ there are five one-magnon bands. One of them
is dispersionless (flat band) with the energy $E_{0}=0$ and this
band belongs to the ground state manifold. The other two flat
bands have energies
\begin{eqnarray}
E_{1} &=&\frac{1-J}{2}  \label{E1} \\
E_{2} &=&\frac{(1+J)^{2}}{2(1-J)}  \label{E2}
\end{eqnarray}

The energies of two more states are%
\begin{equation}
E_{3,4}(\mathbf{k})=\frac{3\left( 1-J\right) }{2}+\frac{J\pm \sqrt{%
(1-J)^{4}+J^{2}-J(1-J)^{2}(\cos k_{x}+\cos k_{y})}}{1-J}  \label{E34}
\end{equation}

The minimal energy of one-magnon excitations is $E_{1}$ for $J\geq
0$ and $E_{2}$ for $J\leq 0$. Note that one of the branches of
$E_{3,4}(\mathbf{k})$ at its minimum touches the lowest flat band
for all values of $J$.

\begin{figure}[tbp]
\includegraphics[width=5in,angle=0]{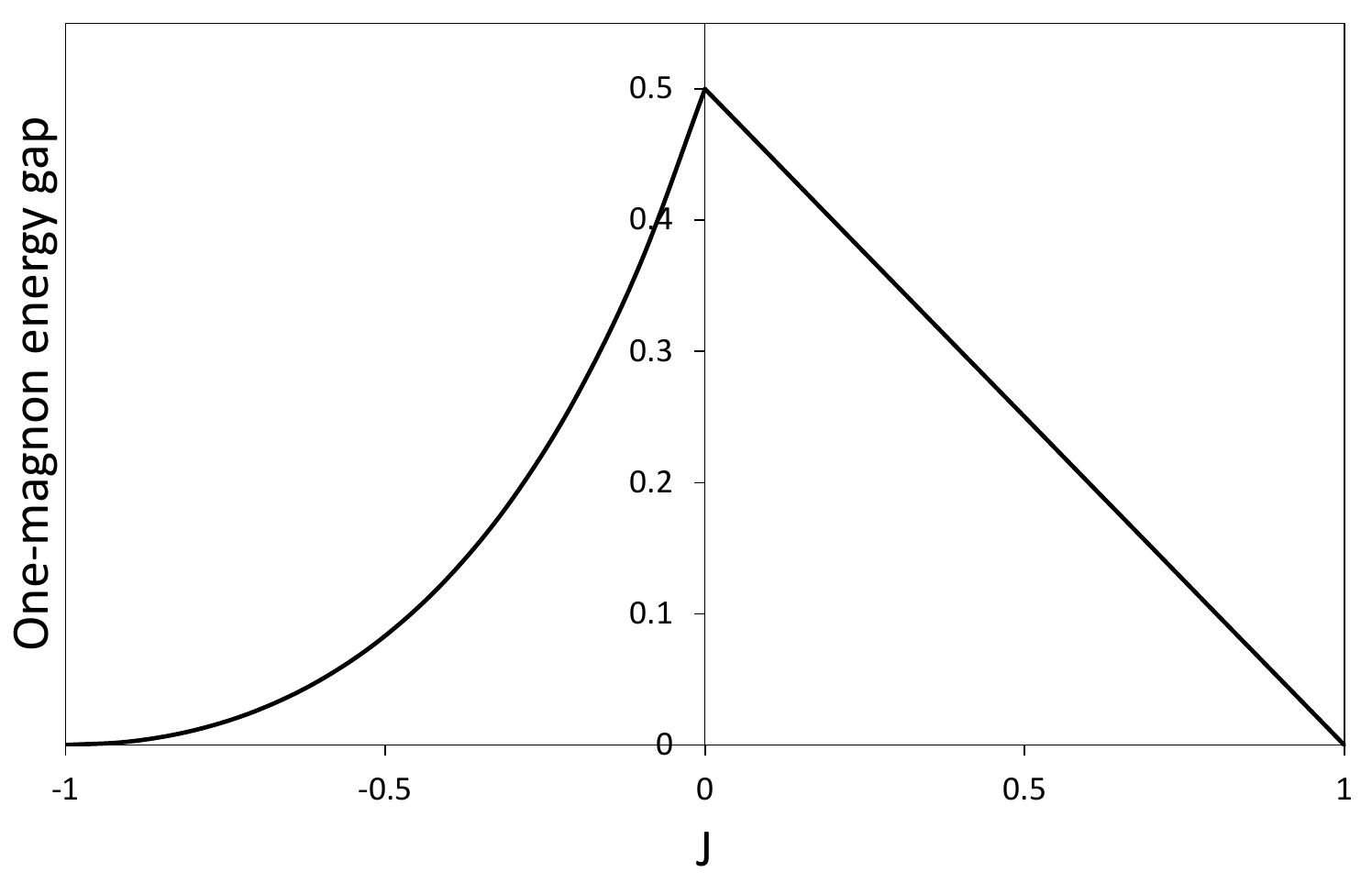}
\caption{The gap in one-magnon excitations of the spin-1/2 F-AF
Heisenberg model on distorted diamond decorated lattice as a
function of interaction $J$.} \label{Fig_gap}
\end{figure}

The dependence of the lowest one-magnon excitation energy on $J$ is shown in
Fig.\ref{Fig_gap}. As can be seen from Fig.\ref{Fig_gap} there is a gap in
the one-magnon spectrum for all values of $J$ in the range $-1<J<1$.

The calculation of the excitation spectrum in other spin sectors
is a rather complicated problem. Numerical calculations are also
limited, since even a $3\times 3$ square lattice contains $45$
spins-$\frac{1}{2}$, and the huge number of ground states
complicates the calculation of the excited state. The calculations
of the total diagonalization for the two-, three-, and four-magnon
spectra on square lattices up to $6\times 6$ show that the
excitation gap in these spin sectors coincides with the gap of one
magnon.

\begin{figure}[tbp]
\includegraphics[width=5in,angle=0]{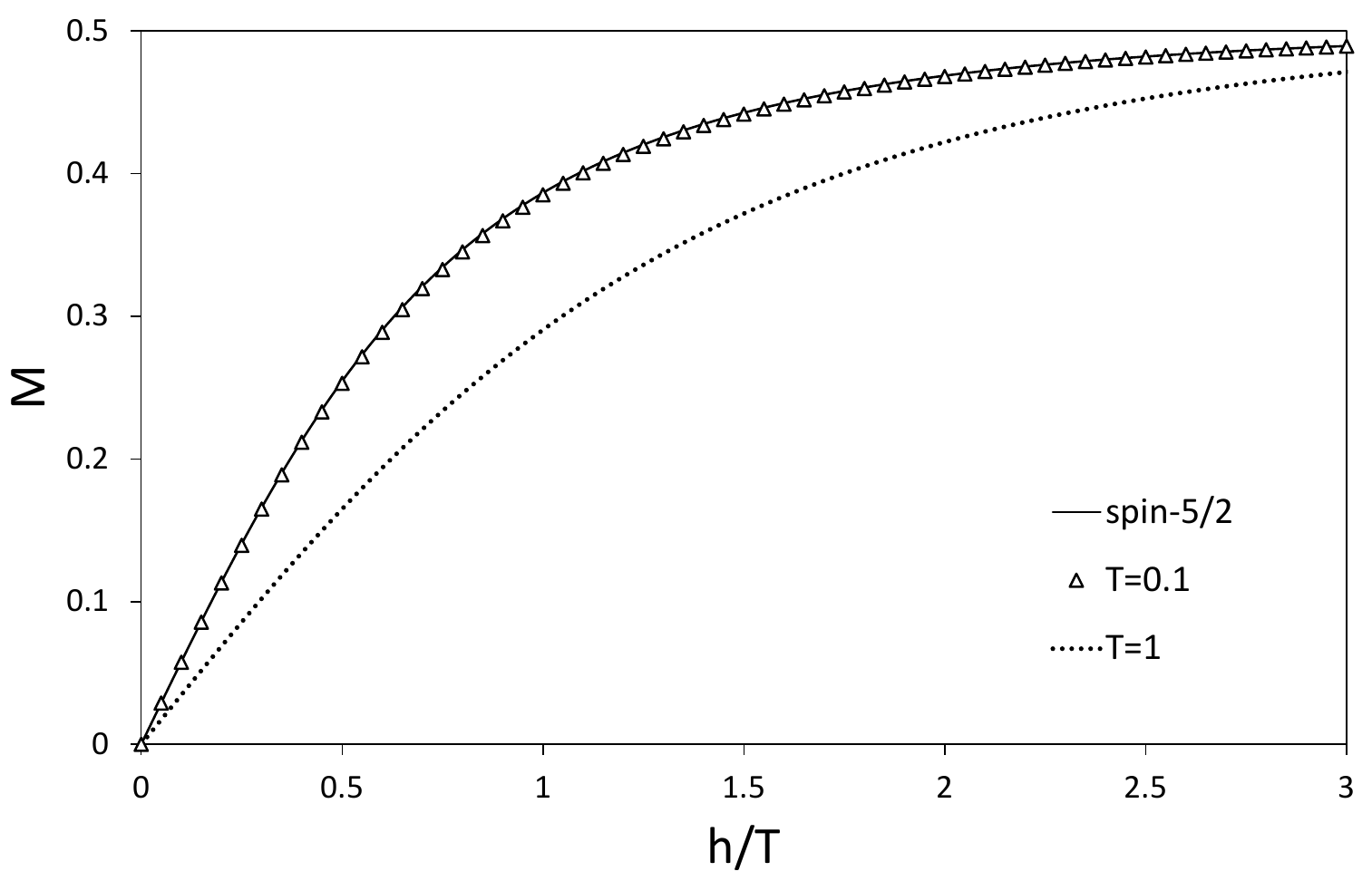}
\caption{Magnetization of the spin-$\frac{1}{2}$ F-AF Heisenberg model on
distorted diamond decorated lattice at $J=0$ for $T=0.1$ (open triangles)
and $T=1$ (dotted line). The magnetization curve of one spin $\frac{5}{2}$
is shown by solid line.}
\label{Fig_M_h}
\end{figure}

Thus, there are three arguments in favor of the fact that the excitation gap
exists for 2D square diamond-decorated lattice: the gap exists for 1D
diamond chain \cite{diamond1d}; the gap exists for the special case $J=0$
and is known exactly $E_{gap}=\frac{1}{2}$; numerical calculations in the
spin sectors with several magnons show the existence of the excitation gap.
Based on the above arguments we expect that there is a gap in the spectrum
for any distorted diamond-decorated lattices, though it can be lower than
the one-magnon gap.

The presence of an excitation gap means that the magnetization and
the magnetic susceptibility at $T<E_{gap}$ are given by the
contributions of the ground state manifold only and they are
identical to those for the system of independent spins
$s=\frac{z+1}{2}$. In particular, the magnetization is $m=0 $ at
$T=0$ and it undergoes a jump from $m=\frac{1}{2}$ in the
ferromagnetic phase to $m=0$ at the critical line. For $T<E_{gap}$
the magnetization per spin $m$ is a function of $\frac{h}{T}$ and,
as an example, for the square lattice it is
$m=\frac{7}{12}\frac{h}{T}$ at $\frac{h}{T}\ll 1$. The
susceptibility $\chi $ diverges as $\chi \sim T^{-1}$. For
illustration we present in Fig.\ref{Fig_M_h} the magnetization
curves for the case $J=0$, where it can be found exactly.

\section{F-AF Heisenberg model on ideal diamond-decorated lattices}

In this section we study the 2D and 3D spin models consisting of
the ideal diamonds, which are the special case $J=-1$ of the
distorted diamonds, studied in Sec.II. In this special case the
exchange interactions on all four sides of the diamond are equal
$J=-1$ and the diagonal interaction is $J_d=1$ (\ref{cond1}). The
Hamiltonian (\ref{Hd}) of the ideal diamond shown in
Fig.\ref{Fig_diamond} takes the form
\begin{equation}
\hat{H}_{\mathbf{i},\mathbf{j}}=-\left( \mathbf{s}_{\mathbf{i}}\mathbf{+s}_{%
\mathbf{j}}\right) \cdot \left( \mathbf{\tau }_{\mathbf{j,i}}\mathbf{+\tau }%
_{\mathbf{i},\mathbf{j}}\right) \mathbf{+\tau }_{\mathbf{i},\mathbf{j}}\cdot
\mathbf{\tau }_{\mathbf{j,i}}+\frac{3}{4}  \label{Hideal}
\end{equation}

As it follows from Hamiltonian (\ref{Hideal}), there is a local
conservation of the composite spin
$\mathbf{L}_{\mathbf{i},\mathbf{j}}=\mathbf{\tau
}_{\mathbf{i},\mathbf{j}}+\mathbf{\tau }_{\mathbf{j,i}}$ on the
diagonals of the diamonds. Composite spin is a conserved quantity
with a quantum spin number $L_{\mathbf{i},\mathbf{j}}=0$ or
$L_{\mathbf{i},\mathbf{j}}=1$, which correspond to the singlet or
triplet state on the diagonal of the diamond, respectively. The
singlet state on the diagonal of the diamond shown in
Fig.\ref{Fig_diamond}, $\left( \tau
_{\mathbf{i},\mathbf{j}}^{+}-\tau _{\mathbf{j,i}}^{+}\right)
\left\vert F\right\rangle $, is an exact state of
$\hat{H}_{\mathbf{i},\mathbf{j}}$, independent of the
configuration of spins $\mathbf{s}_{\mathbf{i}}$ and
$\mathbf{s}_{\mathbf{j}}$.

The number of ground states for ideal diamond models
(\ref{Hideal}) differs from that for distorted diamond models. For
example, if we consider one-magnon sector, then for the distorted
diamond models the number of ground states is equal to the number
of central spins $N$, where the one-magnon states (\ref{phi_1})
are localized. But for ideal diamond models the one-magnon
functions are localized on the diagonal of diamonds forming
singlets, so that the total number of states is equal to the
number of diamonds $N_{b}=\frac{z}{2}N$, in a given lattice. For
1D chain the number of central spins is equal to the number of
diamonds. But for all 2D or 3D lattices, the number of diamonds is
higher than the number of sites. Therefore the ground state
degeneracy of ideal diamond models is greater than the degeneracy
of distorted diamond models in one-magnon sector. As will be shown
below the total ground state degeneracy of ideal diamond models is
greater than the total degeneracy of distorted diamond models.

\subsection{Ground state degeneracy}

Each singlet located on the diagonal of diamond (see
Fig.\ref{Fig_diamond}) effectively breaks the bond between spins
$\mathbf{s}_{\mathbf{i}}$ and $\mathbf{s}_{\mathbf{j}}$, because
in this case $\mathbf{L}_{\mathbf{i},\mathbf{j}}=0$. Therefore,
for ideal diamond models, it is more convenient to calculate the
ground state degeneracy not by sectors with different numbers of
magnons (or the total $S^{z}$), but by sorting through all
possible configurations with different distributions of diamonds
with singlets on the diagonals. This means that all the exact
states of the total Hamiltonian (\ref{H0}) are identified by the
configuration of diamonds with singlets on diagonal, and to
calculate the total ground state degeneracy, it is necessary to
calculate the degeneracy for each configuration of singlets
diagonals and then sum the degeneracy over all configurations. In
this respect the problem of the calculation of the ground state
degeneracy is similar to the \textit{bond percolation problem},
where connected bonds correspond to the diamonds with triplet
diagonal and disconnected bonds to the diamonds with singlet
diagonal.

For a particular configuration of $K$ triplet diagonals (connected
bonds) $\omega _{K}$, the lattice is effectively decompose on,
generally, many non-connected clusters. The ground state of each
of these clusters is the ferromagnetic state with all possible
projections $S^{z}$. Therefore, if the $i$-th cluster contains
$n_{i}$ central spins and $l_{i}$ connected bonds, the total
number of spins in this cluster is $(n_{i}+2l_{i})$ and the ground
state degeneracy of this cluster is $(n_{i}+2l_{i}+1)$. The total
number of ground states for configuration $\omega _{K}$ is the
product of the numbers of ground states of all clusters
\begin{equation}
W(\omega _{K},N)=\prod_{i\in \omega _{K}}(n_{i}+2l_{i}+1)  \label{W}
\end{equation}

\begin{figure}[tbp]
\includegraphics[width=5in,angle=0]{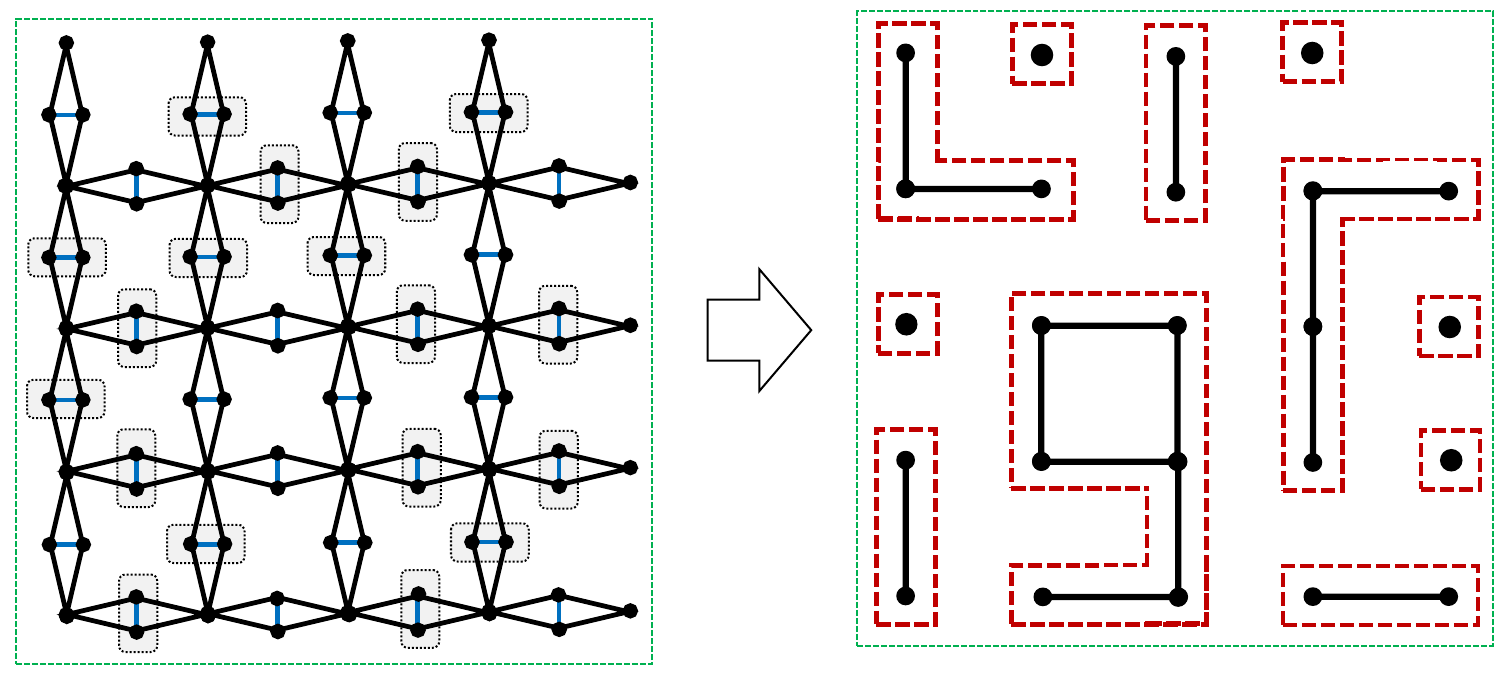}
\caption{Ideal diamond decorated square lattice 4x4 with particular
configuration of diamonds with singlets on diagonal (shaded diagonals) and
the corresponding percolation configuration with connected and disconnected
bonds.}
\label{Fig_percolation}
\end{figure}

As an example, let us consider the square lattice $4\times 4$
($N=16$) with open boundary conditions. One particular
configuration of diamonds with singlet diagonal and the
corresponding connected and disconnected bonds configuration is
shown in Fig.\ref{Fig_percolation}. The shown configuration
contains eleven ferromagnetic clusters: five clusters with one
spin, three clusters with four spins (isolated diamonds), and one
cluster each with $7,10,18$ spins. Therefore, the number of states
of this particular configuration is (\ref{W}): $W=2^5\cdot
5^3\cdot 8\cdot 11\cdot 19=6688000$ states.

\begin{figure}[tbp]
\includegraphics[width=5in,angle=0]{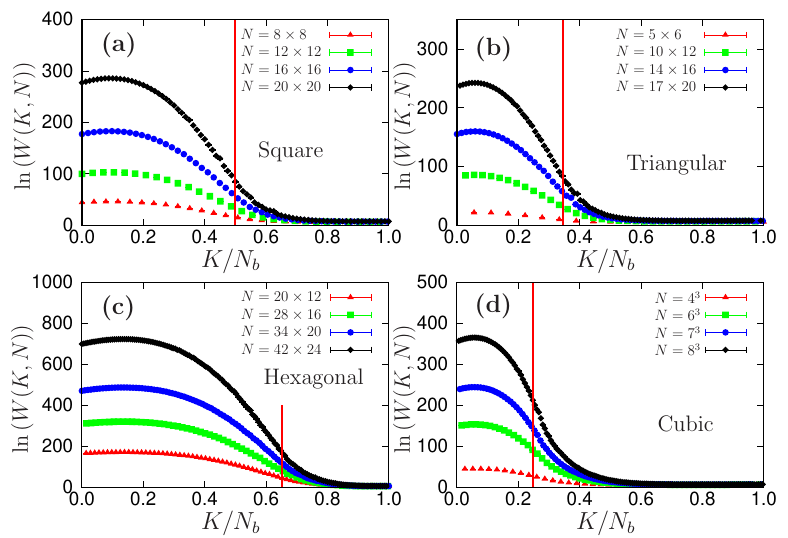}
\caption{Average number of ground states $W(K,N)$ given by
Eq.(\ref{WKN}) plotted as a function of fraction of connected
bonds $K/N_{b}$ for different system size $N$ and for different
lattices: (a) square; (b) triangular; (c) hexagonal; (d) cubic.
The red vertical lines denote the percolation threshold for the
corresponding lattice.} \label{fig:W}
\end{figure}

The number of configurations $\omega _{K}$ with fixed number of
connected bonds $K$ is given by the binomial coefficient
$C_{N_{b}}^{K}=\frac{N_{b}!}{(N_{b}-K)!K!}$. It is convenient to
introduce the average number of
configurations with fixed number of connected bonds $K$%
\begin{equation}
W(K,N)=\frac{1}{C_{N_{b}}^{K}}\sum_{\omega _{K}}W(\omega _{K},N)  \label{WKN}
\end{equation}

In Fig.\ref{fig:W}(a)--(d), we plot the results for $W(K,N)$
versus the fraction of connected bonds $p=K/N_{b}$
($N_{b}=\frac{1}{2}zN$ is the total number of bonds in the
system.). From Fig.\ref{fig:W}(a)--(d), it is evident that the
average number of ground states initially increases smoothly,
reaching a maximum near $p\simeq 0.1$, before decreasing for all
lattices. The characteristic value of $p$ at which $W(K,N)$
rapidly decreases to its minimum is close to the percolation
threshold $p_{c}$ for each respective lattice, as indicated by the
red lines. (The values of the percolation threshold for the
studied lattices are given in Table 2.) This behavior can be
understood as follows: below the percolation threshold
($p<p_{c}$), the system comprises numerous small ferromagnetic
clusters, whose number scales proportionally with the system size.
Consequently, the value of $W$ grows exponentially as $W\sim \exp
(const\cdot N)$ (see Eq.(\ref{W})). Conversely, above the
percolation threshold ($p>p_{c}$), most small ferromagnetic
clusters merge into a single infinite ferromagnetic cluster,
yielding a ground state degeneracy scaling linearly with system
size, i.e., $W\sim N$.

To calculate the total number of ground states $Z(K,N)$ with fixed number of
connected bonds $K$ on the lattice with $N$ sites (central spins) one should
sum up $W(\omega _{K},N)$ for all possible configurations $\omega _{K}$
\begin{equation}
Z(K,N)=\sum_{\omega _{K}}W(\omega _{K},N)=C_{N_{b}}^{K}W(K,N)  \label{ZK}
\end{equation}

\begin{figure}[tbp]
\includegraphics[width=5in,angle=0]{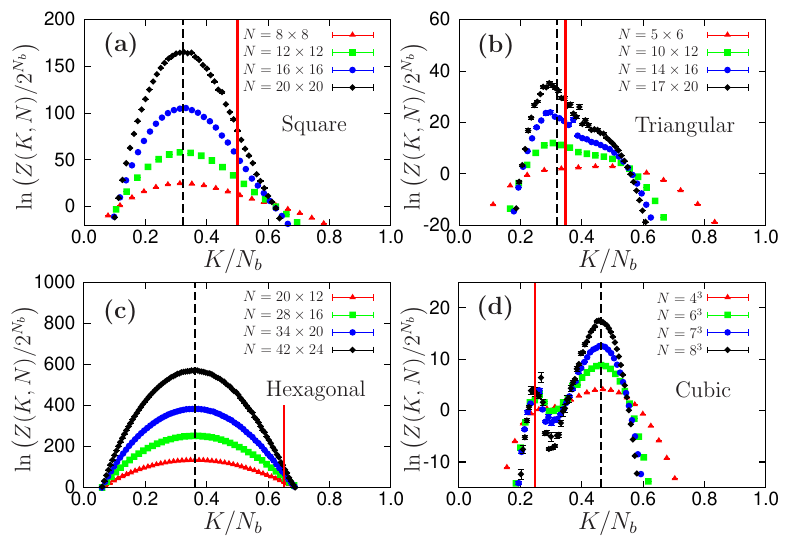}
\caption{The contributions to partition function $Z(K,N)$
Eq.(\ref{ZK}) as a function of the fraction of connected bonds
$K/N_{b}$ and different number of spins $N$ for different
lattices: (a) square; (b) triangular; (c) hexagonal; (d) cubic. }
\label{fig:ZB}
\end{figure}

As follows from Eq.(\ref{ZK}), $Z(K,N)$ is a convolution of
$C_{N_{b}}^{K}$ and $W(K,N)$. While the binomial coefficient
$C_{N_{b}}^{K}$ peaks prominently at $K=N_{b}/2$
($p=\frac{1}{2}$), $W(K,N)$ diminishes as a function of $K$,
shifting the maximum of $Z(K,N)$ toward lower values of $p$
($p<\frac{1}{2}$). This trend is clearly visible in
Fig.\ref{fig:ZB}(a)--(d), where $Z(K,N)$ is plotted against the
fraction of connected bonds $p=K/N_{b}$ for various lattices. As
seen in Fig.\ref{fig:ZB}(a)--(d), the maximal contributions of
$Z(K,N)$ occur approximately at $p_{0}\simeq 0.361$ (hexagonal),
$p_{0}\simeq 0.323$ (square), $p_{0}\simeq 0.32$ (triangular),
$p_{0}\simeq 0.46$ (cubic) lattices. The second, lower maximum of
$Z(K,N)$, which is visible near the percolation threshold $p\simeq
0.25$ for a cubic lattice, is a consequence of exponentially large
values of $W(K,N)$ below the percolation threshold.

Then, the total number of ground states is given by the sum over all
possible numbers of connected bonds $K$
\begin{equation}
Z(N)=\sum_{K=0}^{N_{b}}Z(K,N)  \label{Z}
\end{equation}

\begin{figure}[tbp]
\includegraphics[width=5in,angle=0]{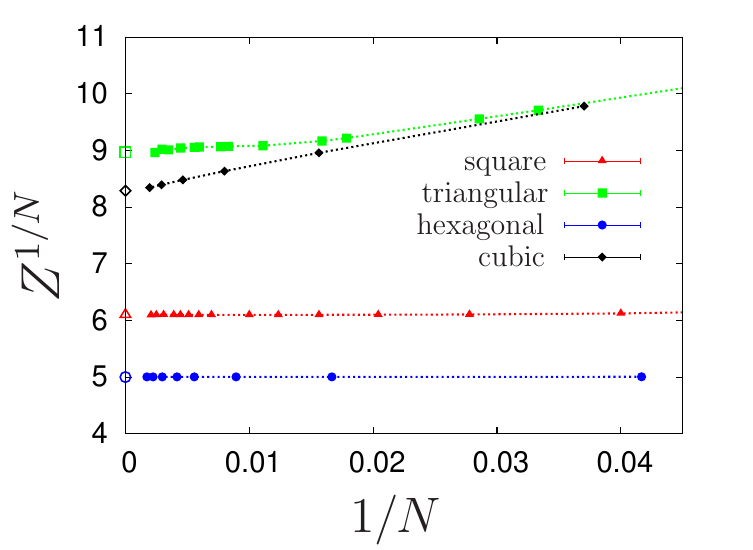}
\caption{The partition function to the power of $1/N$, $Z^{1/N}$,
as a function of $1/N$ for the square, triangular, honeycomb and
cubic lattices.} \label{fig:zn}
\end{figure}

Details of numerical computation of $Z(N)$ are provided in
Appendix B. The obtained results show that for all studied
lattices (hexagonal, square, triangular, cubic) the ground state
degeneracy grows exponentially with $N$, $Z(N)\sim C^{N}$, but
with different $C$. The results for $Z(N)^{1/N}$ vs. $1/N$ for
different lattices are plotted in Fig.~\ref{fig:zn}.

As it is seen in Fig.\ref{fig:zn}, the values $Z(N)^{1/N}$ tend to finite
limits at $N\to \infty $ for all lattices, which gives the thermodynamic
values of constant $C$. The values of $C$ for all studied lattices are
presented in Table 1 along with the values of the exponents for distorted
diamond models on the corresponding lattices.

\begin{table}[tbp]
\caption{The values of exponents for ground state degeneracy for distorted
diamonds and ideal diamonds for different lattices}%
\begin{ruledtabular}
\begin{tabular}{ccc}
Lattice & Distorted  & Ideal    \\
\hline
Chain & $4$ & $4$ \\
Hexagonal & $5$ & $5.00(1)$ \\
Square & $6$ & $6.10(3)$ \\
Triangular  & $8$ & $8.98(3)$ \\
Cubic &  $8$ & $8.26(3)$ \\
\end{tabular}
\end{ruledtabular}
\end{table}

The results in Table 1 demonstrate that for all 2D and 3D lattices
the ground state degeneracy for models with ideal diamonds is
exponentially greater than that the degeneracy for distorted
diamond models (we believe that for a hexagonal lattice, the
exponent of $C$ slightly exceeds the value of 5, although the
accuracy of the results obtained is insufficient to strictly
confirm this). Moreover, the difference increases as the
coordination number of lattice, $z$, increases. However, the
difference in the residual entropy for distorted and ideal models
does not exceed 5\%.

\subsection{Ground state magnetization}

As was established in Sec.II, the ground state magnetization of
the model with distorted diamonds is zero on the critical line. In
this subsection, we study the magnetization for models with ideal
diamonds. The ground state (or zero-temperature) magnetization $M$
for the case when the ground state is macroscopically degenerate
can be calculated as:
\begin{equation}
M^{2}=\frac{1}{Z}\sum_{k=1}^{Z}\left\langle \psi _{k}\right\vert \mathbf{S}%
_{tot}^{2}\left\vert \psi _{k}\right\rangle  \label{M}
\end{equation}%
where $\mathbf{S}_{tot}=\sum \mathbf{s}_{i}$ is total spin
operator of the system and the averaging occurs over all $Z$
ground states $\left\vert \psi _{k}\right\rangle $. The
magnetization defined according to Eq.(\ref{M}) for infinite
lattices is reduced to a long-range order $\left\langle
\mathbf{s}_{\mathbf{i}}\cdot \mathbf{s}_{\mathbf{j}}\right\rangle
$ ($|\mathbf{i}-\mathbf{j}|\to\infty $), averaged over all ground
states. For the pure ferromagnetic systems with $\mathcal{N}$
spins-$\frac{1}{2}$ on any lattice Eq.(\ref{M}) gives the ground
state magnetization per spin
$m=\frac{M}{\mathcal{N}}=\frac{1}{2}$.

Now let us consider a particular configuration with $K$ connected
bonds $\omega _{K}$, which contain a number of non-connected
ferromagnetic clusters. The effective spin of $i$-th cluster
containing $n_{i}$ sites and $l_{i}$ connected bonds is
$S_{i}=\frac{1}{2}n_{i}+l_{i}$. The total number of ground states
$\left\vert \psi _{k}\right\rangle $ for the configuration $\omega
_{K}$ is given by Eq.(\ref{W}). However, the total
$\mathbf{S}_{tot}^{2}$ is the same for all these $W$ states and it
reduces to the sum of $\mathbf{S}_{i}^{2}$ for all clusters
\begin{equation}
\left\langle \psi _{k}\right\vert \mathbf{S}_{tot}^{2}\left\vert
\psi _{k}\right\rangle =\left\langle \psi _{k}\right\vert
\sum_{i\in \omega _{K}}\mathbf{S}_{i}^{2}\left\vert \psi
_{k}\right\rangle  \label{S2}
\end{equation}

Eq.(\ref{S2}) is valid because all clusters are independent, so
that $\left\langle \psi _{k}\right\vert \mathbf{S}_{i}\cdot
\mathbf{S}_{j}\left\vert \psi _{k}\right\rangle =0$ for all
$\left\vert \psi _{k}\right\rangle $ if $i\neq j$. Then, the
magnetization per spin for the configuration $\omega _{K}$ is
given by
\begin{equation}
m^{2}(\omega _{K},N)=\frac{1}{\mathcal{N}^{2}}\sum_{i\in \omega
_{K}}S_{i}(S_{i}+1)  \label{mK}
\end{equation}%
where $\mathcal{N}=N+zN$ is the total number of spins in the lattice with $N$
sites (central spins).

The magnetization per spin, averaged over all ground state manifold, takes
the form
\begin{equation}
m^{2}(N)=\frac{\sum_{K=0}^{N_{b}}\sum_{\omega _{K}}W(\omega
_{K},N)m^{2}(\omega _{K},N)}{\sum_{K=0}^{N_{b}}\sum_{\omega _{K}}W(\omega
_{K},N)}  \label{m}
\end{equation}

It follows from Eq.(\ref{mK}), that if for large system ($N\gg 1$)
the configuration $\omega _{K}$ contains only small clusters
$S_{i}\sim 1$ (the number of which is proportional to $N$), then
the magnetization per spin is very small $m\sim N^{-1/2}$. But if
the configuration $\omega _{K}$ contains the infinite cluster of
the weight $P$ with $S_{i}=P\mathcal{N}$, the magnetization per
spin is finite and equal to the weight of the infinite cluster,
$m=P$. It is known from percolation theory that an infinite
cluster is absent below the percolation threshold
($p=\frac{K}{N_{b}}<p_{c}$), which is different for different
lattices. Above the percolation threshold ($p>p_{c}$) the infinite
cluster rapidly grows $P\sim \left( p-p_{c}\right) ^{\beta }$ with
critical exponent $\beta \simeq 0.14$ for 2D and $\beta \simeq
0.4$ for 3D lattices \cite{percolation_review,beta04_percolation}.
Therefore, the presence or absence of magnetization depends on the
relation between the percolation threshold value $p_{c}$ for a
given lattice and the position of maximum $p_{0}=K_{\max }/N_{b}$
of the function $W(K,N)$, given by Eq.(\ref{ZK}) and shown in
Fig.\ref{fig:ZB}(a)-(d) for different lattices.

\begin{figure}[tbp]
\includegraphics[width=5in,angle=0]{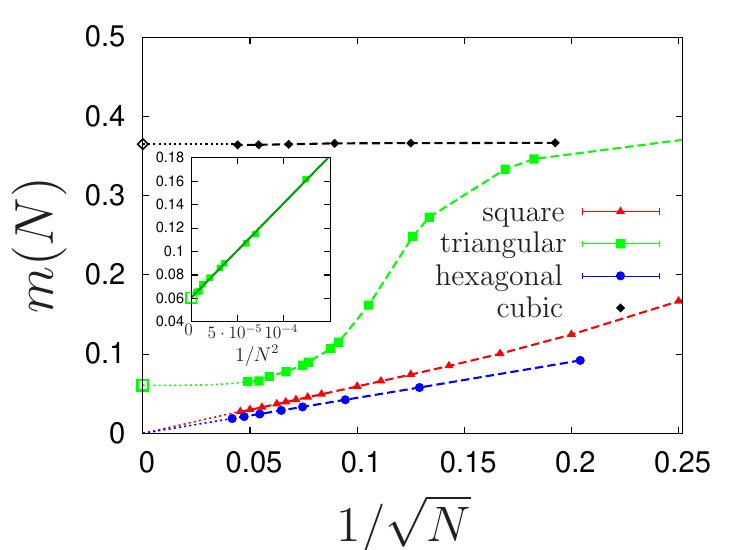}
\caption{Magnetization per spin calculated in accordance with
Eq.(\ref{m}) as a function of $N^{-1/2}$ for square, honeycomb and
cubic lattices. Magnetization for triangular lattice is shown in
the inset versus $N^{-2}$.} \label{fig:M}
\end{figure}

The ground state magnetization $m(N)$ calculated by Eq.(\ref{m})
for different lattices is shown in Fig.\ref{fig:M}. As it is seen
in Fig.\ref{fig:M} the ground state magnetization goes to zero at
$N\to\infty $ for hexagonal and square lattices. This fact is in
accord with that percolation threshold for hexagonal $p_{c}\simeq
0.653$ and for square $p_{c}=0.5$ lattices
\cite{tri_hex_percolation} is significantly greater than the
position of maximum of the function $W(K,N)$, $p_{0}\simeq 0.361$
(hexagonal) and $p_{0}\simeq 0.323$ (square). Therefore, for these
lattices the relative weight of configurations having the infinite
percolation cluster (with $p>p_{c}$) is negligible at $N\to\infty
$. For triangular lattice, the position of maximum of function
$W(K,N)$, $p_{0}\simeq 0.32$, is close to the percolation
threshold $p_{c}\simeq 0.347$ \cite{tri_hex_percolation}, which
leads to a finite, albeit small, magnetization of $m\simeq 0.06$.
For cubic lattice, the position of maximum of $W(K,N)$,
$p_{0}\simeq 0.46$, is much higher than the percolation threshold
$p_{c}\simeq 0.249$ \cite{cubic_percolation}, which means that the
vast majority of ground states have infinite cluster and and this
leads to a high magnetization value $m\simeq 0.365$. All these
results are summarized in the Table 2.

\begin{table}[tbp]
\caption{The values of percolation threshold
\cite{tri_hex_percolation,cubic_percolation}, position of maximum
of function $W(K,N)$ and ground state magnetization per spin for
different lattices.}
\label{Table2}%
\begin{ruledtabular}
\begin{tabular}{cccc}
Lattice & Percolation threshold & Position of maximum of $W(K,N)$ &  Magnetization \\
\hline
Hexagonal & $0.653$ & $0.36$ & $0$ \\
Square & $0.5$ & $0.32$ & $0$ \\
Triangular  & $0.347$ & $0.32$ & $0.06$ \\
Cubic &  $0.249$ & $0.46$ & $0.365$ \\
\end{tabular}
\end{ruledtabular}
\end{table}

\subsection{Excitation spectrum}

The one-magnon spectrum of the ideal diamond decorated square lattice can be
obtained from the one-magnon spectrum of the corresponding distorted diamond
model by setting $J=-1$ in Eqs.(\ref{E1})-(\ref{E34}). This gives two flat
bands with ground state energy $E_{0}=0$, and one flat band with $E_{1}=1$.
The energies of two more states (\ref{E34}) are%
\begin{equation}
E_{3,4}(\mathbf{k})=\frac{5}{2}\left( 1\pm \sqrt{1-\frac{4}{25}(2-\cos
k_{x}-\cos k_{y})}\right)
\end{equation}

Near $k_{x}=k_{y}=0$ the lowest of these two branches reduces to

\begin{equation}
E_{3}(\mathbf{k})=\frac{1}{10}\mathbf{k}^{2}
\end{equation}

This means that the spectrum of the ideal diamond decorated square
lattice is gapless, and the excitation gap on finite systems goes
down as $N^{-1}$ or faster at $N\gg 1$. Similar calculations of
one-magnon spectrum on other lattices show that the one-magnon
excitation gap is $\Delta
E_1=\frac{z}{4(z+1)}(\frac{2\pi}{N})^{2/d}$ ($d$ is dimension of
the lattice) and the ideal diamond models have gapless spectrum
$E(\mathbf{k})\sim \mathbf{k}^{2}$. Thus, the excitation spectrum
differs significantly in models with distorted diamonds and ideal
diamonds.

In the previous subsection, we obtained the non-zero ground state
magnetization for triangular and cubic lattices. Now a natural
question arises: how do gapless excitations affect the
magnetization at a finite temperature $T$. Unfortunately, the
numerical calculations of diamond decorated models on triangular
and cubic lattices are limited by the too small size of the
systems that can be calculated to estimate the thermodynamic
limit. We assume that because the one-magnon spectrum
$E(\mathbf{k})\sim \mathbf{k}^{2}$ is similar to that for the
ferromagnetic models, the behavior of the magnetization at finite
temperatures is also similar to the ferromagnetic model on the
corresponding lattice. Thus, the finite ground state magnetization
for ideal diamond decorated model on triangular lattice is smeared
out by thermal fluctuations at any non-zero temperature, in
accordance with Mermin-Wagner theorem. On the contrary, for cubic
lattice, the magnetization can remain finite for low temperatures,
below the corresponding Curie temperature.

The models with ideal diamond units (\ref{Hideal}) describe the
transition point, at which the condition (\ref{cond1}) of ground
state degeneracy is fulfilled. Now let's look at the ground state
on both sides of this transition point. When $J_d<1$ the ground
state is ferromagnetic. On the other side of the transition point,
when $J_d>1$, the ground state consists of singlets on diagonals
of all ideal diamonds in the system, and free independent central
spins, the so-called monomer-dimer phase. Therefore, the ground
state for $J_d>1$ is $2^N$ degenerate, including the states with
different values of the total spin of the system from $S_{tot}=0$
to $S_{tot}=\frac{1}{2}N$, which is less than the maximum value of
the total spin $S_{\max}=\frac{1+z}{2}N$. Excitations in this
phase are associated with the destruction of singlets on diamond
diagonals and, therefore, are gapped $\Delta E=J_d-1$. This means
that the studied ideal diamond models describes the first-order
phase transition point between the ferromagnetic and the
macroscopically degenerate ground states.

\section{Summary}

In this study, we investigate the ground state properties of the
spin-$\frac{1}{2}$ Heisenberg model on two- and three-dimensional
diamond-decorated lattices. The exchange interactions are chosen
in such a way that the systems are located on the boundary
(critical line) of the ferromagnetic phase. Two types of diamond
units are explored: distorted and ideal, which is a special
symmetrical case of the distorted diamond.

For models with distorted diamonds, the critical line separates
the ferromagnetic phase from the singlet phase. The most
remarkable feature of these models is the existence, along with
conventional localized one-magnon states, of exact multi-magnon
states, localized in one trapping cell. For example, the trapping
cell of a diamond-decorated square lattice can contain up to five
magnons. All these localized multi-magnon states form the ground
states manifold in zero magnetic field. A key result is that the
ground state manifold for a lattice with coordination number $z$
is equivalent to that of non-interacting spins-$\frac{z+1}{2}$,
holding true across all spin sectors $S^{z}$. This equivalence
leads to a macroscopic ground state degeneracy $W=(z+2)^{N}$ ($N$
is a number of trapping cells), resulting in a residual entropy
per spin $\mathcal{S}_{0}=\ln (z+2)/(z+1)$, which reaches $51\%$
of its maximum possible value ($\ln 2$) for the diamond-decorated
square lattice ($z=4$). The ground state magnetization of system
of non-interacting spins is zero, which means that a magnetization
jump from $m=\frac{1}{2}$ in the ferromagnetic phase to $m=0$
occurs at the transition line, indicating magnetic disorder at
finite temperatures.

The spectrum of excitations of the model with the distorted
diamond units is gapped. At $T<E_{gap}$ the magnetic properties of
this model is determined by the ground state manifold, which is
equivalent to the system of non-interacting spins. In particular,
the magnetocaloric effect in the distorted diamond models is
similar to that for paramagnetic salts, which are standard
materials for low-temperature magnetic cooling. For $T<E_{gap}$
the entropy $\mathcal{S}$ depends on the magnetic field $h$ and
the temperature $T$ in the form of the ratio $h/T$, so that during
adiabatic demagnetization the temperature $T$ decreases linearly
with decreasing $h$, so that $T\to 0$ at $h\to 0$. Systems with
higher density of fluctuating spins exhibit a faster cooling rate
$(dT/dh)_{\mathcal{S}}$. From this point of view, the diamond
systems under consideration can be used as the basis of materials
for low-temperature cooling.

Models with macroscopic ground state degeneracy based on distorted diamond
units can be generalized for any lattice and even for any graph, consisting
of distorted diamonds and ferromagnetic bonds. This type of model can also
be generalized for different values of spins forming distorted diamonds and
for anisotropic exchange interactions between spins.

The ground state manifold of models with ideal diamond units can
be represented as randomly distributed ferromagnetic clusters of
variable size and shape surrounded by diamond with singlets on
diagonals, which effectively isolate the ferromagnetic clusters
from each other. The ground state degeneracy is obtained by
numerical calculations, which are similar to those used for the
bond percolation problem. Numerical studies reveal that the ground
state degeneracy exponentially exceeds that of distorted diamond
models. The excitation spectrum of models with the ideal diamonds
is gapless, following a quadratic one-magnon dispersion relation
$E(\mathbf{k})\sim \mathbf{k}^{2}$, analogous to that for the
conventional ferromagnetic models. The ground state magnetic
ordering depends on dimension and coordination number of the
lattice: finite magnetization persists in 3D cubic and 2D
triangular lattices at $T=0$, and is absent in 2D square/hexagonal
systems. These facts suggest a potential magnetic ordering at
finite temperatures for 3D models with ideal diamond units.

Our findings demonstrate that both distorted and ideal diamond-based models
provide an intriguing insights into ground state properties and could serve
as valuable platforms for studying complex magnetic phenomena and designing
materials for advanced applications, including low-temperature cooling
technologies.

The numerical calculations were carried out with use of the ALPS libraries
\cite{alps}.

\appendix

\section{Proof of the existence of localized multimagnon states}

In this appendix we prove that the localized multi-magnon states
presented in Eq.(\ref{phi_k}) are exact ground states of
Hamiltonian (\ref{H0}). To do this, we need to prove that the
states (\ref{phi_k}) are exact ground states of each of four
diamonds adjacent to the star $\mathbf{i}$. Since all these
diamonds are equivalent, it suffices to prove this for any of
these diamond Hamiltonians, let it be
$\hat{H}_{\mathbf{i},\mathbf{j}}$. For this purpose, we will
extract the operator $\sigma _{\mathbf{i},\mathbf{j}}^{+}$ that
relates to the chosen diamond ($\mathbf{i},\mathbf{j}$) from the
sum in Eq.(\ref{phi_k}) and rewrite it in the form
\begin{equation}
\sum_{\left\{ \mathbf{r}_{1}\ldots \mathbf{r}_{k}\right\} }\sigma _{\mathbf{i%
},\mathbf{r}_{1}}^{+}\ldots \sigma _{\mathbf{i},\mathbf{r}_{k}}^{+}=\sigma _{%
\mathbf{i},\mathbf{j}}^{+}\Phi _{\mathbf{i},\mathbf{j}}^{(k-1)}+\Phi _{%
\mathbf{i},\mathbf{j}}^{(k)}  \label{A1}
\end{equation}%
with%
\begin{equation}
\Phi _{\mathbf{i},\mathbf{j}}^{(k)}=\sum_{\left\{ \mathbf{r}_{1}\ldots
\mathbf{r}_{k}\right\} }^{\prime }\sigma _{\mathbf{i},\mathbf{r}%
_{1}}^{+}\ldots \sigma _{\mathbf{i},\mathbf{r}_{k}}^{+}  \label{A2}
\end{equation}%
where the prime above the sum sign means that there are no configurations
with the factor $\sigma _{\mathbf{i},\mathbf{j}}^{+}$ in the sum.

After the transformation (\ref{A1}) the multi-magnon states (\ref{phi_k})
takes the form
\begin{equation}
\hat{\varphi}_{\mathbf{i}}^{(k)}=\Phi _{\mathbf{i},\mathbf{j}}^{(k)}+\left(
s_{\mathbf{i}}^{+}+\sigma _{\mathbf{i},\mathbf{j}}^{+}\right) \Phi _{\mathbf{%
i},\mathbf{j} }^{(k-1)}+s_{\mathbf{i}}^{+}\sigma _{\mathbf{i},\mathbf{j}%
}^{+}\Phi _{\mathbf{i},\mathbf{j}}^{(k-2)}  \label{A3}
\end{equation}

The first term in Eq.(\ref{A3}) corresponds to the state of the
diamond ($\mathbf{i},\mathbf{j}$) with all spin down, which is
exact ground state of $\hat{H}_{\mathbf{i},\mathbf{j}}$. The
second and the third terms in Eq.(\ref{A3}) correspond to the
states (\ref{1}) and (\ref{2}) of the diamond ($
\mathbf{i},\mathbf{j}$), which are also exact ground states of
$\hat{H}_{\mathbf{i},\mathbf{j}}$ as it was mentioned in Section
2.

\section{Numerical simulation of the ground state degeneracy}

\label{app:MC}

For computation of the average number of configurations $W(K,N)$
(\ref{WKN}) and the partition function $Z(K,N)$ shown in
Figs.\ref{fig:W},~\ref{fig:ZB}, we used Monte Carlo method. First,
a random realization $\omega _{K}$ with $K$ connected bonds is
created. Then, the realization $\omega _{K}$ split into a set of
connected clusters using the
Hoshen-Koplman~algorithm~\cite{Hoshen76}. The calculation of the
number of sites $n_{i}$ and connected bonds $l_{i}$ for each
cluster $i$ of realization $\omega _{K}$ allows us to compute the
statistical weight $W(\omega _{K},N)$ using Eq.(\ref{W}).
Repeating the above steps for $N_{MC}$ random realizations $\omega
_{j,K}$ with fixed $K$ connected bonds, we calculate the average
value of $W(\omega _{K},N)$
\begin{equation}
W(K,N)=\frac{1}{N_{MC}}\sum_{j=1}^{N_{MC}}W(\omega _{j,K},N)
\label{B1}
\end{equation}

Then, the partition function $Z(K,N)$ is calculated using
Eq.(\ref{ZK}).

In order to prepare the data for Figs.\ref{fig:W},~\ref{fig:ZB} we
used the averaging over $N_{MC}=10^{7}$ random configurations for
each value of $0\leq K\leq N_{b}$. This set of configurations was
divided into 10 series for evaluation of numerical inaccuracy.

The total partition function $Z(N)$ can be computed directly by
summing $Z(K,N)$ over all possible values of $K$, as described in
Eq.(\ref{Z}). However, this approach is computationally
inefficient. Instead, we employ a more efficient algorithm that
samples configurations closer to the maximum of $Z(K,N)$ (see
Fig.\ref{fig:ZB}) more frequently, taking into account the proper
normalization. Specifically, we adopt the following procedure.
First, we select a probability $p$ for a bond to be connected,
implying a probability of $1-p$ for it to be disconnected. Each
bond configuration $\omega _{i}$ is then created by connecting
bonds probabilistically according to $p$ or leaving them
disconnected with probability $1-p$. Thus, the probability of
generating a configuration with exactly $K$ connected bonds is
given by $p^{K}(1-p)^{N_{b}-K}$. Therefore, the contribution of
the configuration with $K$ connected bonds should be divided by
$p^{K}(1-p)^{N_{b}-K}$. Since the majority of configurations
produced by this method typically have $K\sim pN_{b}$, we set $p$
equal to the location of the maximum of $Z(K,N)$ for each lattice,
$p_0=K_{\max}/N_b$. Then, the partition function is calculated as
a result of averaging over $N_{MC}$ random bond configurations
$\omega _{i}$, created by this algorithm:
\begin{equation}
Z_{MC}(N)=\frac{2^{N_{b}}}{N_{MC}}\sum_{i=0}^{N_{MC}}\frac{W(\omega
_{i},N)}{p_0^{K(\omega _{i})}(1-p_0)^{N_{b}-K(\omega _{i})}},
\label{B2}
\end{equation}%
where $K(\omega _{i})$ is the number of connected bonds in the
bond configuration $\omega _{i}$, $W(\omega _{i},N)$ is the
statistical weight of the configuration $\omega _{i}$ in
accordance with Eq.(\ref{W}). The factor $2^{N_b}$ comes from the
total number of all terms in $Z(N)$.

The square of magnetization is computed in the same way
\begin{equation}
m_{MC}^{2}(N)=\frac{2^{N_{b}}}{N_{MC}}\sum_{i=0}^{N_{MC}}\frac{%
m^{2}(\omega _{i},N)W(\omega _{i},N)}{p_0^{\mathcal{K}(\omega
_{i})}(1-p_0)^{N_{b}-\mathcal{K}(\omega _{i})}}, \label{B3}
\end{equation}%
where the square of magnetization $m^{2}(\omega _{i},N)$ of
configuration $\omega _{i}$ is given by Eq.(\ref{mK}).

Since the majority of configurations generated by this algorithm
have $K$ values close to $K_{\max}$, the number of configurations
far from $K_{\max}$ is negligible, especially for large lattices,
where certain $K$-sectors may even be entirely absent in the sums
(\ref{B2}) and (\ref{B3}). To validate the correctness of the
algorithm, we selected a uniformly distributed set of
probabilities $\{p_{j}\}$ and performed averaging of the final
result over this entire set. We confirmed that all methods yield
consistent results. Nevertheless, for a fixed number $N_{MC}$ of
random configurations, the most precise result is obtained when
using a single point $p=p_0$, positioned at the maximum of
$Z(K,N)$. Table \ref{Table2} lists the values of $p_0$ for
different lattices, denoted as "Position of maximum". Throughout
our calculations, we employed $N_{MC}=10^{10}$ bond configurations
for each data point in Figs.\ref{fig:zn} and ~\ref{fig:M}.

\bibliography{diamond2D}

\end{document}